\documentclass[%
 aip,
 amsmath,amssymb,
 reprint,%
]{revtex4-1}

\usepackage{graphicx}
\usepackage{dcolumn}
\usepackage{bm}

\usepackage[utf8]{inputenc}
\usepackage[T1]{fontenc}
\usepackage{mathptmx}
\usepackage{color}

\usepackage{caption}
\usepackage{subcaption}

\usepackage[page]{appendix}

\begin{document}

\preprint{AIP/123-QED}

\title[]{
Excited State Mean-Field Theory without Automatic Differentiation
}

\author{Luning Zhao}
\affiliation{
Department of Chemistry, University of Washington, Seattle, Washington 98195, USA 
}
\author{Eric Neuscamman}
 \email{eneuscamman@berkeley.edu.}
\affiliation{
Department of Chemistry, University of California, Berkeley, California 94720, USA 
}
\affiliation{Chemical Sciences Division, Lawrence Berkeley National Laboratory, Berkeley, CA, 94720, USA}

\date{\today}

\begin{abstract}
We present a formulation of excited state mean-field theory in which the derivatives
with respect to the wave function parameters needed for wave function optimization
(not to be confused with nuclear derivatives) are expressed analytically
in terms of a collection of Fock-like matrices.
By avoiding the use of automatic differentiation and grouping Fock builds together,
we find that the number of times we must access the memory-intensive two-electron integrals
can be greatly reduced.
Furthermore, the new formulation allows the theory to exploit existing strategies for
efficient Fock matrix construction.
We demonstrate this advantage explicitly via the shell-pair screening strategy,
with which we achieve a cubic overall cost scaling.
Using this more efficient implementation, we also examine the theory's ability
to predict charge redistribution during charge transfer excitations.
Using coupled cluster as a benchmark, we find that by
capturing orbital relaxation effects and avoiding
self-interaction errors, excited state mean field theory out-performs other low-cost
methods when predicting the charge density changes of charge transfer excitations.
\end{abstract}

\maketitle

\section{Introduction}
\label{sec:intro}

Whether one is talking about light harvesting, \cite{Grondelle2011,Neaton2017}
photo-catalysts, \cite{Feldmann2018,Domen2019}
core spectroscopy, \cite{Oosterbaan2018,vidal2019new,zheng2019}
metal-to-ligand charge transfer,\cite{siefermann2014,CT2018}
or non-adiabatic dynamics,\cite{MCTDH,Tully2012}
reliable predictions of excited state properties are extremely valuable.
However, the leading theoretical methods for modeling excited states
are fundamentally more approximate than their ground state counterparts
due to their reliance on additional approximations.
Linear response (LR) methods, for example, assume that the excited state is in some
way close to the ground state in state space, leading in practice to
a situation in which crucial orbital relaxation effects
\cite{Ziegler2009,Subotnik2011,Ziegler2015,Zhao2019}
are simply absent in
configuration interaction singles (CIS) \cite{Dreuw2005}
and time-dependent density functional theory (TDDFT) \cite{Dreuw2005,Hirata1999,Burke2004}
and only treated in a limited manner in singles and doubles
equation-of-motion Coupled Cluster (EOM-CCSD) theory. \cite{Krylov2008}
These shortcomings contribute to the difficulties that CIS and TDDFT
have with charge transfer (CT) states \cite{Subotnik2011,Zhao2019}
and to the eV-sized errors that EOM-CCSD often makes for doubly excited states.
\cite{watts1996,Zhao2016}
As these difficulties arise in part from a lack of orbital relaxation
following an excitation, the development of excited state methods that
have fully-relaxed, excited-state-specific orbitals is strongly desirable.

Towards this end, the recently-introduced excited state mean-field (ESMF)
theory \cite{Shea2018} attempts to provide a minimally-correlated reference
state for single excitations in which the orbitals are fully relaxed.
In some aspects, ESMF is similar to the $\Delta$SCF approach
\cite{bagus1965scf,Pitzer1976scf,argen1991xray,Gill2009dscf}
and the restricted open-shell Kohn Sham (ROKS) approach, \cite{Shaik1999,Kowalczyk2013}
but in contrast to these methods it delivers a completely spin-pure
wave function by construction and can handle states in which multiple
singly-excited configurations are present in a superposition.
ESMF is also similar to the restricted active space self-consistent 
field (RASSCF) method \cite{olsen1988rasscf,malmqvist1990restricted}
since both methods optimize orbitals for a 
set of selected configurations.
However, these two approaches have important differences.
While RASSCF typically approaches excited states via 
state-averaging, ESMF is a wholly excited-state-specific method.
In addition, while ESMF avoids the need to define an active space,
this simplicity comes at a price: it is not appropriate for
strongly correlated settings such as conical intersections.
Like Hartree-Fock (HF) theory, \cite{Szabo-Ostland,MolElecStruc} the energetic accuracy
of ESMF itself is limited due to missing correlation effects, which previous work
has shown to cause a bias towards underestimating excitation energies. \cite{Shea2019}
Nonetheless, ESMF acts as a powerful starting point for perturbation theory
\cite{Shea2019} and can be used to construct a novel form of density functional theory. \cite{Zhao2019}
The perturbation theory results are particularly exciting, with preliminary testing
showing an accuracy competitive with EOM-CCSD. \cite{Shea2019}
Like $\sigma$-SCF \cite{vanVoorhis2017sigmaSCF,Voorhis2019}
and some quantum Monte Carlo approaches to excited states,
\cite{umrigar1988optimized,Zhao2016,Shea2017,pineda2019}
ESMF takes a variational approach to orbital relaxation in which it minimizes
a function whose global minimum is the desired excited state.
However, unlike these approaches, the generalized variational principle (GVP)
that ESMF minimizes is defined in terms of the norm of the energy gradient
with respect to wave function parameters such as CI coefficients and orbital
rotation matrices,\cite{Shea2019}
and so the optimization requires some information about second derivatives
of the energy with respect to the wave function variables.
On the bright side, automatic differentiation (AD) via TensorFlow
\cite{tensorflow2015-whitepaper}
or any other general-purpose AD framework can be employed to evaluate
the necessary derivatives while maintaining the same Fock-matrix-build
cost-scaling of an ESMF energy evaluation.
On a less positive note, this approach requires the memory-intensive
two-electron integrals (TEIs) to be accessed many times during each evaluation
of the GVP's gradient and makes it difficult to take advantage of
acceleration techniques like shell-pair screening, \cite{Ochsenfeld2005}
density fitting, \cite{Head-Gordon2006}
and tensor hyper contraction. \cite{THC1,THC2,THC3}
Thus, while AD has been extremely helpful in quickly developing an initial
implementation for testing ESMF theory (see Appendix for examples of
cumbersome expressions that it allows one to avoid),
it severely limits the practical efficiency of the approach.
In this study, we develop explicit analytic expressions for the objective function's
gradient with respect to wave function variables that simplify into a collection
of nine Fock-like matrix builds
that can be carried out together (thus minimizing the number of times the TEIs need
to be accessed) and which can benefit straightforwardly from acceleration methods.
The result is a dramatic speedup compared to our previous TensorFlow implementation,
thus allowing ESMF theory to be used in significantly larger systems.
Now, to avoid confusion between the analytic gradients we are discussing with
other types of analytic gradients, let us be very clear:  there are no
nuclear gradients in this study.
All of the gradients we investigate are with respect to wave function variables such
as configuration coefficients or orbital rotation parameters.

We will begin with a brief review of ESMF theory, including the
wave function ansatz and the GVP employed in its optimization.
We will then discuss why the TensorFlow implementation 
of the energy derivatives is inefficient,
after which we derive the analytic energy derivative expressions
and show how they can be formulated using Fock-builds.
Following these theoretical developments, we test the efficiency in two
charge-transfer systems and demonstrate a speedup of two orders of magnitude
relative to our previous implementation.
We then turn to an investigation of charge transfer in a
minimally-solvated environment that would not have been possible
with the previous AD-based approach to evaluating the GVP derivatives.
Encouraged by the results, we round out our results section with additional
testing of charge density changes in four other charge transfer examples.
Finally, we conclude with a summary and some
comments on future directions.

\section{Theory}
\label{sec:theory}

\subsection{ESMF Method}
In ESMF, the wave function ansatz is written as
\begin{equation}
    \label{eqn:esmf_wfn}
    \left|\Psi\right>=e^{\hat{X}}\left(c_0\left|\Phi\right>+\sum_{ia}{\sigma_{ia}\hat{a}^{\dagger}_{a\alpha}\hat{a}_{i\alpha}\left|\Phi\right>+\tau_{ia}\hat{a}^{\dagger}_{a\beta}\hat{a}_{i\beta}\left|\Phi\right>}\right)
\end{equation}
in which $\left|\Phi\right>$ is the restricted Hartree-Fock determinant, and the coefficients 
$\sigma_{ia}$ and $\tau_{ia}$ correspond to excitations of an alpha-spin and a beta-spin electron, 
from the $i$th occupied orbital to the $a$th virtual orbital. The operator $\hat{X}$ is defined as,
\begin{equation}
    \label{eqn:x_ope}
    \hat{X}=\sum_{pq}{X_{pq}\hat{a}_p^{\dagger}\hat{a}_q}
\end{equation}
in which $X$ is an anti-symmetric matrix so that the orbital rotation operator
$\hat{U}=\mathrm{exp}(-\hat{X})$ is unitary. 
Note that this orbital rotation will be state specific, which means that
different states' wave functions will not be strictly orthogonal.
While this trait has not prevented ESMF from acting as a powerful reference for
perturbation theory \cite{Shea2019} or from accurately predicting charge
density changes (see Section \ref{sec:density_changes}), it can cause
some concern, and so we have for the sake of completeness included the
overlaps between ground and excited states for the systems we study
in the Supplemental Materials.
Here, we simply note that these overlaps are quite small.

Although there is an argument to be made that a single open-shell configuration
state function should be seen as the minimally correlated reference function
for excited states, we choose to include all singly-excited configurations in
order to handle states in which two or more of these configurations exist
in a superposition, as for example occurs in the low-energy spectrum of N$_2$.
Although such states are technically multi-configurational, they certainly
do not require a general-purpose strongly correlated ansatz and indeed are
already treated at a qualitatively correct level by CIS.
That said, the open shell character of ESMF (in which two electrons correlate
their positions so as to not reside in the same spatial orbital simultaneously)
does involve more correlation than HF theory and so it is a step further away
from a true mean-field theory in which no correlation is present at all.

In the initial development of ESMF theory, \cite{Shea2018}
the ansatz was optimized using the Lagrangian-based objective function,
\begin{equation}
    \label{eqn:esmf_lag_multi}
    L_{\vec{\lambda}}=W+\vec{\lambda}\cdot\nabla E
\end{equation}
in which $W$ was seen as an approximated excited state variational principle
whose purpose is to guide the optimization to the energy stationary point
associated with a particular excited state and
$\vec{\lambda}$ is a set of Lagrange multipliers that ensure that the optimized 
wave function is indeed an energy stationary point. 
The approximate variational principle was chosen as
\begin{equation}
    \label{eqn:esmf_es_var}
    W=(\omega-E)^2\approx\frac{\left<\Psi\right|(\omega-H)^2\left|\Psi\right>}{\left<\Psi|\Psi\right>}
\end{equation}
which, although successful in initial testing on small molecules,
was found to have multiple shortcomings.
First, the target function is not bounded from below with respect to Lagrangian 
multipliers, making simple quasi-Newton methods difficult to use directly.
Instead, the expression $|\nabla L_{\vec{\lambda}}|^2$ was minimized,
which further increased the computational cost by necessitating an additional
layer of automatic differentiation (still the right cost scaling, but now containing
components that formally involve triple derivatives of the energy).
Further, a more recent study \cite{Shea2019} found that this approach
can show poor numerical conditioning, sometimes requiring hundreds of quasi-Newton 
iterations to converge. 

In order to address these two problems, a finite-difference
Newton-Raphson (NR) method was developed for an objective function based
not on Lagrange multipliers but instead on a GVP. \cite{Shea2019}
\begin{equation}
    \label{eqn:esmf_gvb}
    L_{\mu\chi}=\chi\left(\mu\left(\omega-E\right)^2+\left(1-\mu\right)|\nabla E|^2\right)+(1-\chi)E
\end{equation}
Starting with $\chi$ set to 1, $\mu$ is gradually reduced to zero during
the optimization so as to ensure convergence to the stationary point with
energy closest to $\omega$.
Unlike the target function in Equation \ref{eqn:esmf_lag_multi}, this approach
is bounded from below, allowing both NR and quasi-Newton methods to be employed without the need for an additional layer of AD.
In addition, one can switch $\chi$ to 0 close to convergence and rely on stationary-point
methods like NR, thus potentially benefiting from even more efficient gradients and
such method's super-linear convergence.
In practice, initial testing has shown this approach to be more robust and more
efficient than Equation \ref{eqn:esmf_lag_multi} in a variety of systems.
\cite{Shea2019}

\subsection{Analytic Derivatives}

Even though this new approach improved the optimization's efficiency,
its implementation still relied on AD for the derivatives of both the energy and
$L_{\mu\chi}$, which, although convenient, leads to an unnecessarily high prefactor
in the method's cost due to frequent access of the TEIs and the handling of the
TEIs as a dense 4-index array without the efficiencies that accelerated Fock-build
methods enjoy.
In order to address this source of inefficiency, we will now derive explicit expressions
for the analytic derivatives of the ESMF energy and objective function and show
that they can be formulated as a set of Fock matrix builds.
We will focus on the special case of singlet excited states,
whose wave function can be written as
\begin{equation}
    \label{eqn:esmf_singlet_wfn}
    \begin{split}
        \left|\Psi\right>&=e^{\hat{X}}\left(c_0\left|\Phi\right>+\sum_{ia}{\sigma_{ia}\left(\hat{a}^{\dagger}_{a\alpha}\hat{a}_{i\alpha}+\hat{a}^{\dagger}_{a\beta}\hat{a}_{i\beta}\right)\left|\Phi\right>}\right) \\
        &=e^{\hat{X}}\left(c_0\left|\Phi\right>+\left|\Psi_{\mathrm{CIS}}\right>\right) \\
    \end{split} 
\end{equation}
in which the alpha and beta electron excitations have the same coefficients and we have grouped 
the linear combination of singly excited determinants to the CIS ($\left|\Psi_{\mathrm{CIS}}\right>$)
wave function. Although the derivation will be based on singlet excited states,
a generalization to the triplet case is straightforward. 

\subsubsection{Notation}

Before we derive ESMF energy, target function, and derivatives, we introduce the notations
used in our derivations.
Orbital index $i$, $j$, $k$ denote occupied orbitals. Index $a$, $b$, $c$ denote
virtual orbitals.
We use the indices $p$, $q$,  $r$, $s$ for general orbitals.
We denote the first $N_o$ rows of the $\textbf{V}$
matrix as $\bm{\Theta}$, where $N_o$ is the total number of occupied orbitals. The last 
$N_v$ rows of $\textbf{V}$ are denoted as the matrix $\bm{\Gamma}$,
where $N_v$ is the number of virtual orbitals.
Similarly, we denote the corresponding blocks of the matrix $\textbf{M}$
(see Table \ref{tab:notations}) as $\textbf{R}$ and $\bm{\Phi}$. 

We define the ``generalized'' Coulomb, exchange, and Fock matrices as,
\begin{equation}
    \label{eqn:J_and_K}
    \begin{split}
            J[\textbf{D}]_{pq}&=\sum_{rs}{D_{rs}(rs|pq)} \\
            K[\textbf{D}]_{pq}&=\sum_{rs}{D_{rs}(pr|qs)} \\
            F[\textbf{D}]_{pq}&=2J[D]_{pq}-K[D]_{pq} \\
    \end{split}
\end{equation}
in which $(rs|pq)$ are the two-electron integrals in the atomic
orbital basis in 1122 order, and $\textbf{D}$ is a
``generalized'' density matrix which is not necessarily symmetric.
For a summary of the notation for the scalar and matrix quantities we use,
as well as for some matrix operations, see Table \ref{tab:notations}. 

\begin{table}
    \caption{Summary of Notation
         \label{tab:notations}
    }
    \begin{tabular}{c c}
    \hline
    Description & Notation \\
    \hline\hline
    One-electron Integrals in AO Basis & $\textbf{G}$ \\
    Two-electron Integrals in AO Basis & $(pq|rs)$ \\
    RHF Orbital Coefficients & $\textbf{C}$ \\
    RHF Determinant Coefficient & $c_0$ \\
    Excited Determinant Coefficients & $\bm{\sigma}$ \\
    ESMF Orbital Coefficients & $\textbf{V}=\textbf{U}^T\textbf{C}$ \\
    $c_0$ Lagrangian Multiplier & $\mu_0$ \\
    $\sigma$ Lagrangian Multiplier & $\bm{\mu}$  \\
    $X$ Lagrangian Multiplier & $\textbf{M}$ \\
    Transformed $U$ Lagrangian Multiplier & $\textbf{W}=\textbf{M}^T\textbf{C}$ \\
    First $N_o$ Rows of $\textbf{V}$ & $\bm{\Theta}$ \\
    Last $N_v$ Rows of $\textbf{V}$ & $\bm{\Gamma}$ \\
    First $N_o$ Rows of $\textbf{W}$ & $\textbf{R}$ \\
    Last $N_v$ Rows of $\textbf{W}$ & $\bm{\Phi}$ \\
    General Fock Matrix & $\textbf{F}[\textbf{D}]$ \\
    Wave Function Square Norm & $N_2=c_0^2+2\sum_{ia}{c_{ia}^2}$ \\
    $\textbf{A}$ Matrix & $\bm{\Gamma}^T\bm{\sigma}^T\bm{\sigma \Gamma}-\bm{\Theta}^T\bm{\sigma\sigma}^T\bm{\Theta}$ \\
    $\textbf{B}$ Matrix & $\bm{\Theta}^T\bm{\sigma}^T\bm{\sigma\Gamma}-\textbf{R}^T\bm{\sigma\sigma}^T\bm{\Theta}$ \\
    Matrix Trace & $\mathrm{Tr}[\textbf{O}]=\sum_{p}{O_{pp}}$ \\\
    Matrix Inner Product & $\textbf{O}\cdot \textbf{V}=\sum_{pq}{O_{pq}V_{pq}}$ \\
    \hline
    \end{tabular}
\end{table}


\subsubsection{ESMF Energy}
The ESMF energy is computed as
\begin{equation}
    \label{eqn:esmf_init_eng}
    \begin{split}
            E&=\left(c_0^2\left<\Phi|e^{-\hat{X}}\hat{H}e^{\hat{X}}|\Phi\right>+2c_0\left<\Phi|e^{-\hat{X}}\hat{H}e^{\hat{X}}|\Psi_{\mathrm{CIS}}\right>\right. \\
            &\left.+\left<\Psi_{\mathrm{CIS}}|e^{-\hat{X}}\hat{H}e^{\hat{X}}|\Psi_{\mathrm{CIS}}\right>\right)/N_2 \\
    \end{split}
\end{equation}

By splitting the Hamiltonian into one- and two-body operators, the energy can be formulated as
\begin{equation}
    \label{eqn:esmf_eng}
    E=\frac{E_1+E_2}{N_2}
\end{equation}

The ESMF energy can now be evaluated in the same formalism as CIS energy with rotated molecular 
orbital coefficients as,
\begin{equation}
    \label{eqn:rot_mo_coeff}
     \mathrm{\textbf{V}}=\textbf{U}^T\textbf{C}
\end{equation}
in which the one- and two-body components are
\begin{equation}
    \label{eqn:esmf_energy_1b}
    \begin{split}
            E_1&=2N_2\mathrm{Tr}[\bm{\Theta}\textbf{G}\bm{\Theta}^T]+4c_0\mathrm{Tr}[\bm{\Theta}\textbf{G}\bm{\Gamma}^T\bm{\sigma}^T] \\
            &+2\mathrm{Tr}[\bm{\sigma\Gamma}\textbf{G}\bm{\Gamma}^T\bm{\sigma}^T-\bm{\Theta} \textbf{G}\bm{\Theta}^T\bm{\sigma\sigma}^T] \\
    \end{split}
\end{equation}
and
\begin{equation}
    \label{eqn:esmf_energy_2b}
    \begin{split}
        E_2&=N_2\textbf{F}[\bm{\Theta}^T\bm{\Theta}]\cdot\left(\bm{\Theta}^T\bm{\Theta}\right)+4c_0\textbf{F}[\bm{\Theta}^T\bm{\Theta}]\cdot\left(\bm{\Theta}^T\bm{\sigma\Gamma}\right) \\
        &+2\textbf{F}[\bm{\Theta}^T\bm{\Theta}]\cdot \textbf{A}+2\textbf{F}[\bm{\Theta}^T\bm{\sigma\Gamma}]\cdot\left(\bm{\Theta}^T\bm{\Theta}\right) \\
    \end{split}
\end{equation}
respectively. The first terms in the one- and two-body energy expression come from the contribution 
of the un-excited determinant with rotated orbitals. The second terms are due to the cross terms 
between excited and un-excited determinants, which are zero in CIS with canonical orbitals due to 
Brillouin theorem. However, these terms are non-zero in ESMF with rotated canonical orbitals. 
The last two terms arise from the contributions of purely excited determinants. 

These expressions reveal that the most expensive part of the ESMF energy is the 
construction of two Fock matrices:
$\textbf{F}[\bm{\Theta}^T\bm{\Theta}]$ and
$\textbf{F}[\bm{\Theta}^T\bm{\sigma\Gamma}]$.
Note that this is different from HF theory, in which only one Fock matrix,
$\textbf{F}[\bm{\Theta}^T\bm{\Theta}]$, is needed to evaluate the energy.
In fact, we note that if we set $\bm{\sigma}=\textbf{0}$ and $\textbf{U}=\textbf{I}$
then the ESMF energy becomes,
\begin{equation}
    \label{eqn:esmf_eng_hf}
    E=2Tr[\textbf{PG}]+F[\textbf{P}]\cdot\textbf{P}
\end{equation}
in which $P_{rs}=\sum_{i}{C_{ri}C_{si}}$. one would immediately 
realize that this is the HF energy expression.

\subsubsection{Derivatives of Lagrangian-based Objective Function}

Starting from the energy, we have derived the first derivatives
$\partial E/\partial c_0$, $\partial E/\partial \bm{\sigma}$,
$\partial E /\partial\bm{\Theta}$, and $\partial E /\partial \bm{\Gamma}$,
detailed expressions for which can be found in the Appendix.
Note that these derivatives require the construction of
one additional Fock matrix, $\textbf{F}[\textbf{A}]$,
in addition to the two required for the energy itself.
In this subsection, we use these components to find the analytic derivatives of the
Lagrangian-based objective function of Equation \ref{eqn:esmf_lag_multi}.
Once we have these derivatives in hand, we will see in the next subsection
how they can be modified to produce analytic derivatives of the
newer GVP-based objective function of Equation \ref{eqn:esmf_gvb}.
Note that, as the derivation of the different variables' derivatives is very
similar, we will often work in terms of derivatives with respect to a generic wave
function variable $x$, which could be either a configuration coefficient
from Equation \ref{eqn:esmf_wfn}
or an element of the $\bm{X}$ matrix from Equation \ref{eqn:x_ope}.

We first expand the Lagrange multiplier dot product as
\begin{equation}
    \label{eqn:lag_mult_long}
        L_{\vec{\lambda}}=W+\mu_0\frac{\partial E}{\partial c_0}+\sum_{ia}{\mu_{ia}\frac{\partial E}{\partial\sigma_{ia}}}+\sum_{pq}{M_{pq}\frac{\partial E}{\partial X_{pq}}}
\end{equation}
and note that the derivative of $W$ with respect to any variable $x$ is simple
once the energy first derivatives have been evaluated.
\begin{equation}
    \label{eqn:w_der} 
    \frac{\partial W}{\partial x}=-2\left(\omega-E\right)\frac{\partial E}{\partial x}
\end{equation}
Noting that the remaining part of $L$ can be written in terms of the auxiliary quantities
$L_1$, $L_2$, and $L_3$ (defined in the Appendix), the 
objective function becomes
\begin{equation}
    \label{eqn:esmf_total_l}
    \begin{split}
            L&=W+\left(\mu_0L_1-2\mu_0Ec_0\right)/N_2 \\
            &+\left(L_2-4E\sum_{ia}{\sigma_{ia}\mu_{ia}}\right)/N_2+L_3/N_2.
    \end{split}
\end{equation}
Inspecting the definition of $L_3$, we find that this approach requires 
two additional Fock builds: $\textbf{F}[\textbf{R}^T\bm{\Theta}]$ and
$\textbf{F}[\textbf{R}^T\bm{\sigma\Gamma}]$,
bringing us to five total Fock builds for this approach to constructing
the energy, its derivatives, and $L$.

Although we again relegate the details to the Appendix,
we find that evaluating the derivatives of $L_1$, $L_2$, and $L_3$
with respect to $c_0$, $\bm{\sigma}$, $\bm{\Theta}$, and $\bm{\Gamma}$
requires another four Fock builds:
$\textbf{F}[\bm{\Theta}^T\bm{\mu\Gamma}]$,
$\textbf{F}[\bm{\Theta}^T\bm{\sigma\Phi}]$, 
$\textbf{F}[\bm{\Theta}^T\bm{\mu\sigma}^T\bm{\Theta}]$, and
$\textbf{F}[\textbf{B}]$.
Thus, after constructing a total of nine Fock-like matrices,
the evaluation of the derivatives of $L$ with respect to all of the ESMF wave function
variables amounts to an inexpensive (relative to the Fock builds)
collection of operations on matrices whose dimensions are no worse than the
number of orbitals.
We have so far derived derivatives with respect to the elements of $\textbf{V}$, 
but the actual orbital rotation parameters are the elements of $\textbf{X}$.
To get all the way to derivatives with respect to $\textbf{X}$, we first transform the 
derivatives with respect to $\textbf{V}$ to the derivatives with respect to $\textbf{U}$
through the following way,
\begin{equation}
    \label{eqn:der_v_to_u}
    \frac{\partial E}{\partial\textbf{U}}=\textbf{C}\frac{\partial E}{\partial\textbf{V}^T}
\end{equation}
In order to get derivatives with respect to $\textbf{X}$,  there are at least two
possible routes.
On the one hand, AD (e.g.\ via TensorFlow) can be used to complete the last step in
a reverse-accumulation approach, in which the values for $\partial E / \partial \bm{U}$
that we have calculated via our analytic expression are fed in to reverse accumulation
through the matrix exponential function.
On the other hand, if we wish to strictly avoid AD (although for this part of the evaluation
there is not such a clear efficiency case for doing so)
we can instead redefine our orbital rotation matrix as
\begin{equation}
    \label{eqn:redefine_U}
    \bm{U} \rightarrow \bm{\tilde{U}} \exp ( -\bm{X} )
\end{equation}
where after each optimization step we reset $\bm{X}$ to zero by absorbing
its rotation into $\bm{\tilde{U}}$ and then exploiting the simple relationship
$\exp(\bm{X})=\bm{1}+\bm{X}$.
Either way, once we have evaluated the energy derivatives with respect to $\bm{U}$,
converting them to derivatives with respect to $\bm{X}$ is inexpensive compared to the
Fock builds.

\subsubsection{Derivatives of GVP-based Objective Function}

We now turn to the derivatives of the GVP-based objective function in
Equation \ref{eqn:esmf_gvb} with respect to some wave function variable $x$.
\begin{equation}
    \label{eqn:esmf_l_gvp_der}
    \begin{split}
        \frac{\partial L_{\mu\chi}}{\partial x}&=\chi\left(-2\mu\left(\omega-E\right)\frac{\partial E}{\partial x}+2\left(1-\mu\right)\sum_{y}{\frac{\partial E}{\partial y}\frac{\partial^2 E}{\partial x\partial y}}\right) \\
        &+\left(1-\chi\right)\frac{\partial E}{\partial x} \\
    \end{split}
\end{equation}
Now, compare this expression to the derivatives of $L_{\vec{\lambda}}$.
\begin{equation}
    \label{eqn:esmf_l_mul_der}
    \begin{split}
        \frac{\partial L_{\vec{\lambda}}}{\partial x}&=-2\left(\omega-W\right)\frac{\partial E}{\partial x}+\sum_{y}{\lambda_y\frac{\partial^2 E}{\partial x\partial y}}
    \end{split}
\end{equation}
If we first evaluate the energy derivatives, which requires three Fock builds, we can replace
the Lagrange multipliers in Eq.\ (\ref{eqn:esmf_l_mul_der}) with these energy
derivatives, at which point the approach discussed above for evaluating the derivatives
of $L_{\vec{\lambda}}$ can be used for evaluating the term inside the large parentheses
in Eq.\ (\ref{eqn:esmf_l_gvp_der}).
Thus, as for the Lagrange multiplier objective function,
the GVP objective function's derivatives can be evaluated at the cost of nine Fock builds
(the three already done for the energy derivatives plus the remaining six).

\subsection{Shell-Pair Screening}

In total, our analytic evaluation of the derivatives needed for ESMF optimization
requires building nine Fock matrices. \\
\begin{tabular}{l l}
   & \\
    \hspace{15mm}
    1. \hspace{1mm} $\textbf{F}[\bm{\Theta}^T\bm{\Theta}]$
      & \hspace{10mm} 6. \hspace{1mm} $\textbf{F}[\bm{\Theta}^T\bm{\mu\Gamma}]$ \\[1mm]
    \hspace{15mm}
    2. \hspace{1mm} $\textbf{F}[\bm{\Theta}^T\bm{\sigma\Gamma}]$
      & \hspace{10mm} 7. \hspace{1mm} $\textbf{F}[\bm{\Theta}^T\bm{\sigma\Phi}]$ \\[1mm]
    \hspace{15mm}
    3. \hspace{1mm} $\textbf{F}[\textbf{A}]$
      & \hspace{10mm} 8. \hspace{1mm} $\textbf{F}[\bm{\Theta}^T\bm{\mu\sigma}^T\bm{\Theta}]$ \\[1mm]
    \hspace{15mm}
    4. \hspace{1mm} $\textbf{F}[\textbf{R}^T\bm{\Theta}]$
      & \hspace{10mm} 9. \hspace{1mm} $\textbf{F}[\textbf{B}]$ \\[1mm]
    \hspace{15mm}
    5. \hspace{1mm} $\textbf{F}[\textbf{R}^T\bm{\sigma\Gamma}]$ &  \\
   & \\
\end{tabular}
\\
Although this ensures that ESMF has the same asymptotic cost scaling as HF theory,
it is clearly going to suffer from a higher prefactor.
How to mitigate this extra cost?
In this study, we employ the shell-pair screening approach to Fock matrix construction,
both because it exploits sparsity in the TEIs in larger systems and because it allows
us to easily group Fock builds together in order to minimize the number of times the
sparse TEIs must be accessed.
%
%
The shell-pair screening algorithm can be summarized as follows.
At the beginning of an ESMF
optimization we loop over all four indices of the two-electron integrals $\left(p,q,r,s\right)$, and 
for each index pair $\left(p,q\right)$, find the $\left(r,s\right)$ index pair that
maximizes the absolute value of integral $|\left(pq|rs\right)|$.
The shell pair $\left(p,q\right)$ is discarded unless $\max_{r,s}|\left(pq|rs\right)|$
is above a certain threshold, which in this work we set to 10$^{-9}$ a.u., small enough so
that we achieve direct agreement with results from our older TensorFlow-based implementation.
After this screening, Fock builds proceed according to Eq.\ (\ref{eqn:J_and_K})
by looping over the retained shell pairs.  
Crucially, multiple Fock matrices can be constructed during a single loop through
the TEIs.
As the TEIs still take up a lot of memory even after screening, it is more cache
efficient to evaluate multiple fock matrices at once.
Thus, when using the Lagrange multiplier objective function, we evaluate
all nine Fock matrices during a single loop over the shell pairs.
For the GVP objective function derivatives, we require two loops through the shell pairs,
the first to evaluate the first three matrices and the second to evaluate the other six,
which in that case depend on the first three due to
the Lagrange multiplier values having been set equal to the energy derivatives.
Finally, note that we have implemented a simple approach to shared-memory parallelism
by threading the loop over shell pairs.


\section{Results}
\label{sec:results}

\subsection{Computational Details}
All of the ESMF results are obtained via our own software, which extracts one- and 
two-electron integrals from PySCF.\cite{Sun2018} The DFT, TDDFT, CIS, ROHF, and CCSD results were 
obtained from QChem.\cite{qchem} We use the VESTA\cite{VESTA} software to plot the density difference 
between ground and excited state. The molecular geometries can be found in the Supplemental Materials. 

\begin{figure}[h!]
\centering
\includegraphics[width=8.5cm,angle=0,scale=1.0]{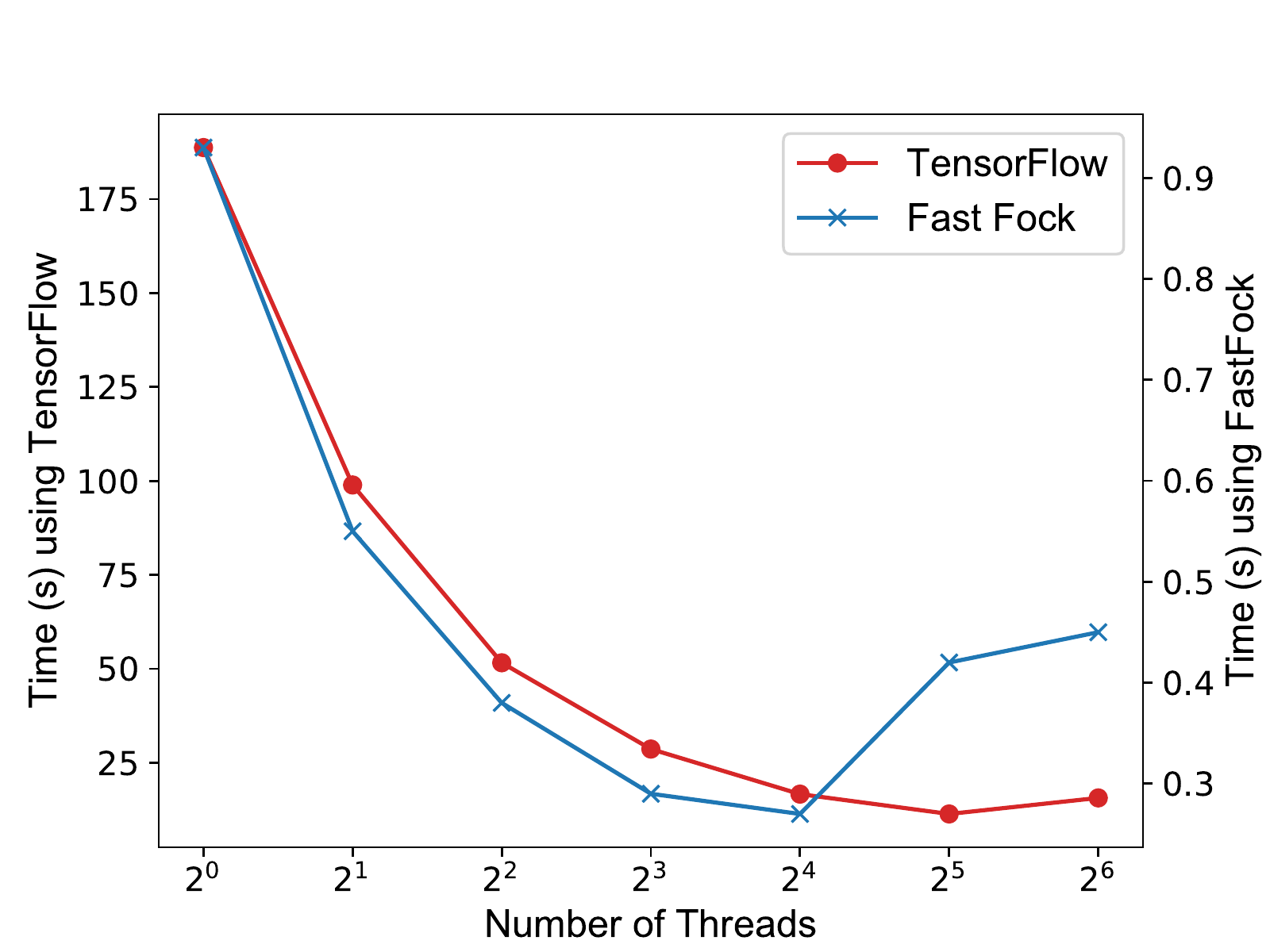}
\caption{
  Time taken to evaluate the 24 Lagrangian gradients required for inverting the Hessian
  during the first finite-difference NR iteration of the Cl$^-$-H$_2$O optimization.
        }
\label{fig:cl_h2o}
\end{figure}

\begin{figure}[h!]
\centering
\includegraphics[width=8.5cm,angle=0,scale=1.0]{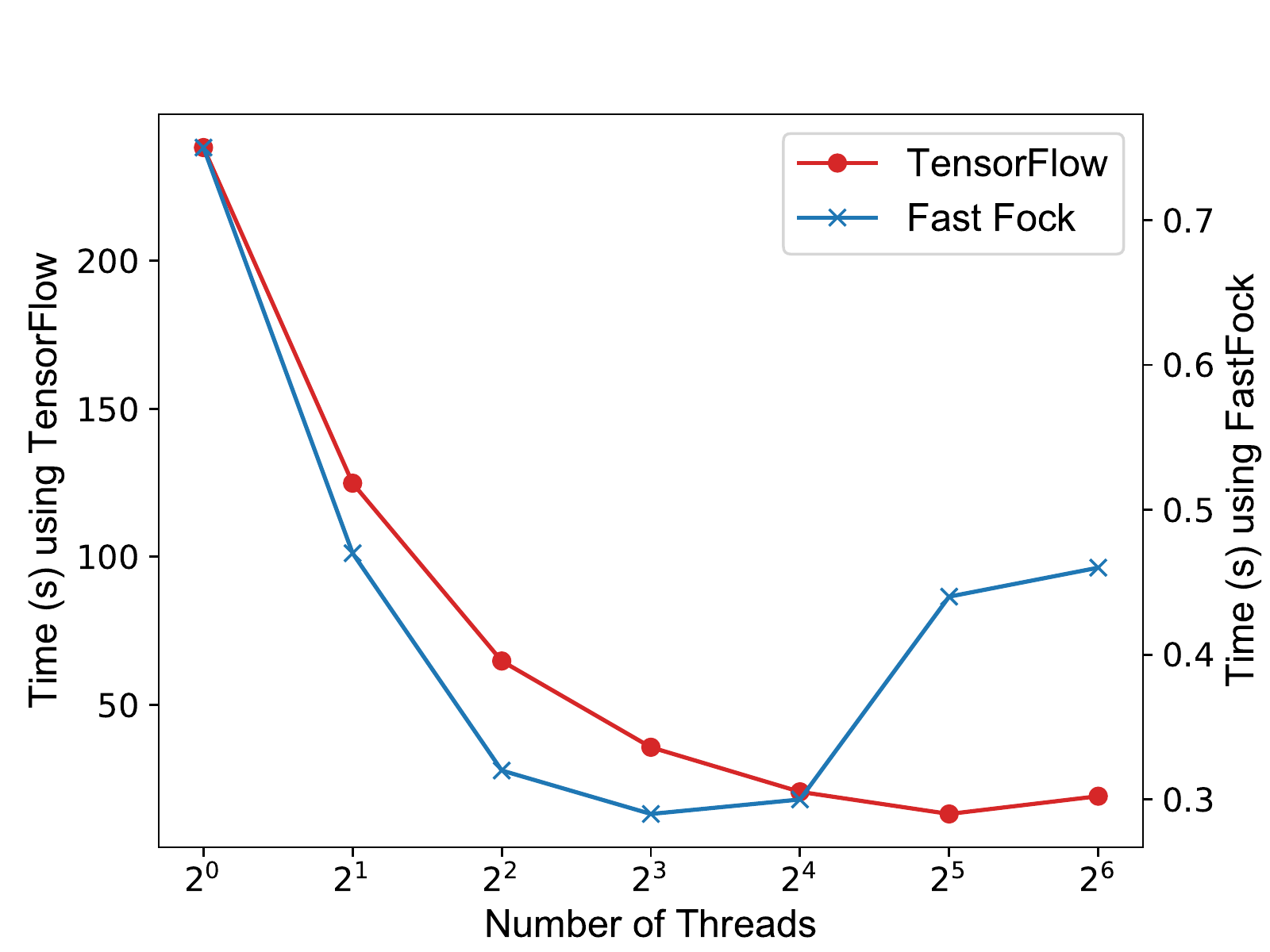}
\caption{
  Time taken to evaluate the 22 Lagrangian gradients required for inverting the Hessian
  during the first finite-difference NR iteration of the NH$_3$-F$_2$ optimization.
        }
\label{fig:nh3_f2}
\end{figure}

\subsection{Efficiency Gain and Cost Scaling}

Here we compare the cost of two example ESMF optimizations when performed with
our group's previous AD-based implementation (denoted here as TensorFlow)
with the new implementation based on analytic expressions and shell-pair screening
(denoted here as FastFock).
Working in the cc-pVDZ basis, \cite{Dunning1988} 
we optimize the lowest singlet charged transfer excited states of
the Cl$^-$-H$_2$O dimer and the NH$_3$-F$_2$ dimer,
both of which have been studied before. \cite{Zhao2006,Zhao2019,Kim2000}
The optimization uses the finite-difference NR approach, \cite{Shea2019}
in which the NR linear equation is solved via the generalized minimal residual
method (GMRES), with the Hessian-vector multiplication evaluated via a simple
finite difference formula involving gradients of the objective function.
In Figures \ref{fig:cl_h2o} and \ref{fig:nh3_f2},
we compare the amount of time that the two implementations took
to solve the linear equation for each optimization's first NR step.
Although we would clearly need to work with larger systems for our simple
multi-threading approach to saturate the 32 cores on the processor,
we see a roughly 100-fold increase in speed in both cases when
running 16 threads (be careful to note the different left and right axes).
When run in serial, the new implementation is 202 and 314 times faster
than the TensorFlow implementation for the two different systems.

We now illustrate the origin of the efficiency of our new ESMF implementation 
by comparing the performance of the 5 different ESMF implementation
options described in Table \ref{tab:implemenattion}. 
These five implementations differ in programming language/framework, 
how they conduct differentiation, 
and whether shell-pair screening is performed. Option 1 is our old TensorFlow ESMF implementation 
and option 5 is our new FastFock implementation. Option 2-4 connect option 1 with option 5 
by changing one variable at a time. We compare the amount of time each implementation 
took to evaluate the same 22 Lagrangian gradients in Figure \ref{fig:nh3_f2}. 

\begin{table}[h]
    \caption{Comparison of how long five different ESMF implementations
             took to evaluate the Lagrangian gradients in Figure \ref{fig:nh3_f2}.
             The ``Fock Build'' column indicates which language the
             bottleneck Fock-build step is implemented in.
         \label{tab:implemenattion}
    }
    \begin{tabular}{c c c c c}
    \hline
    Option & Differentation & Fock Build & Shell-Pair Screening & Time (s) \\
    \hline\hline
    1 & Automatic & TensorFlow & No & 35.64\\
    2 & Analytic & TensorFlow & No & 3.41\\
    3 & Analytic & NumPy & No & 3.69\\
    4 & Analytic & C & No & 0.33\\
    5 & Analytic & C & Yes & 0.35\\
    \hline
    \end{tabular}
\end{table}

As shown in the Table, by merely changing automatic differentiation to analytic 
differentiation but maintaining the TensorFlow framework, we already see a
10-fold speed-up in terms of the gradient evaluation. Such an acceleration shows the 
benefits of deriving analytic derivatives and collecting all the Fock-like 
matrices, since doing so allows us to drastically reduce the number of times
we must access the memory-intensive two-electron integrals. Another 10-fold
speed-up is achieved by moving the Fock build implementation from Python to C.
In the final row, we see that this system is too small to benefit from
shell-pair screening, and so we now turn our attention to larger cases in
which this benefit can be realized.

\begin{figure}[b]
\centering
\includegraphics[width=8.5cm,angle=0,scale=1.0]{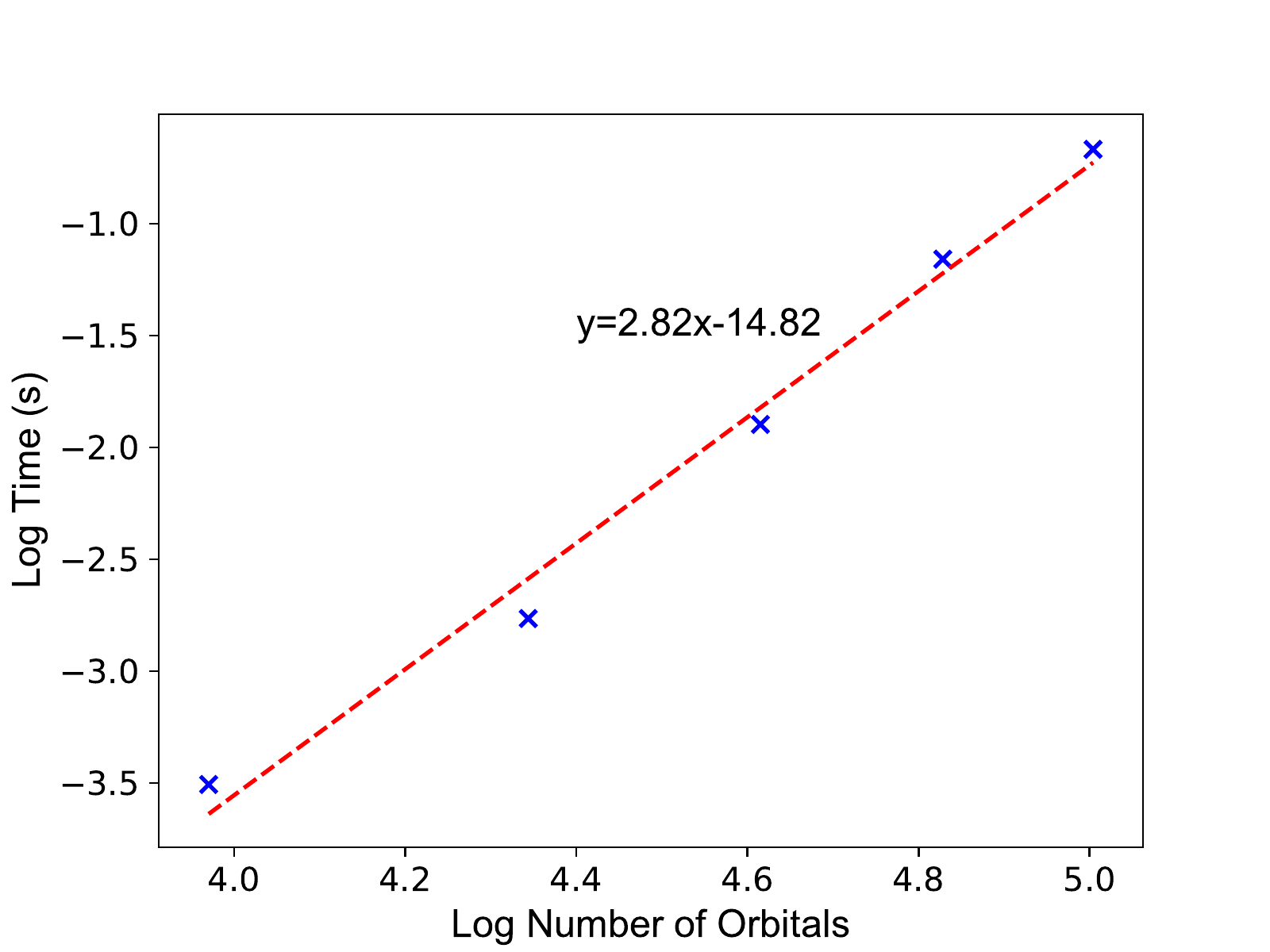}
\caption{
  A log-log plot of the time taken to evaluate one Lagrangian gradient as
  a function of the number of atomic orbitals for the NH$_3$-(H$_2$O)$_n$ system
  with varying numbers of water molecules.  8 OpenMP threads were used.
        }
\label{fig:nh3_h2o}
\end{figure}

In order to show the cost scaling of our method, in Figure \ref{fig:nh3_h2o} we plot the 
time taken to evaluate one Lagrangian gradient as a function of the number of orbitals in 
a NH$_3$-(H$_2$O)$_n$ system. The structure is obtained by first putting water molecules 
around NH$_3$ randomly and then performing geometry optimizations in a 6-31G$^\ast$ basis 
with MP2. 
As more water molecules are added, we find that the implementation's cost
scaling is roughly $O(N^3)$.
This is expected, as although the shell-pair screening can greatly reduce
the cost of the Fock builds,
the remaining matrix-matrix multiplications involved in forming our
derivatives should give an $N^3$ scaling.


\begin{table}[t]
\caption{Excitation energies (eV) for the TensorFlow and FastFock implementations,
         demonstrating that the shell-pair screening within the latter has not
         affected the energetics.
         \label{tab:ee_compare}
}
\begin{tabular}{c c c}
\hline\hline
System & TensorFlow & Fast Fock \\
\hline
Cl$^-$-H$_2$O & 4.7195 & 4.7195 \\
NH$_3$-F$_2$ & 4.5367 & 4.5367 \\
\hline\hline
\vspace{2mm}
\end{tabular}
\end{table}

\begin{figure}[b]
\centering
\includegraphics[width=8.5cm,angle=0,scale=0.70]{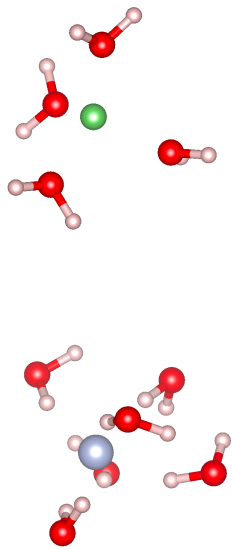}
\caption{The geometry of the minimally-solvated charge transfer system, in which a Li atom
         surrounded by four water molecules can be seen above a F atom surrounded by six water
         molecules.
         The geometry was determined by two MP2 optimizations in the
         6-31G$^{\ast}$ basis, \cite{Petersson88}
         one for the positively-charged Li cluster in isolation and one for the
         negatively-charged F cluster in isolation.
         The two clusters were then arranged one atop the other such that
         the Li and F atoms were 8.05$\mathring{A}$ apart.
         See Supplemental Materials for the atomic coordinates.
        }
\label{fig:lif_geo}
\end{figure}

\begin{figure*}
    \centering
    \begin{subfigure}{0.3\textwidth}
        \centering
        \includegraphics[scale=0.30]{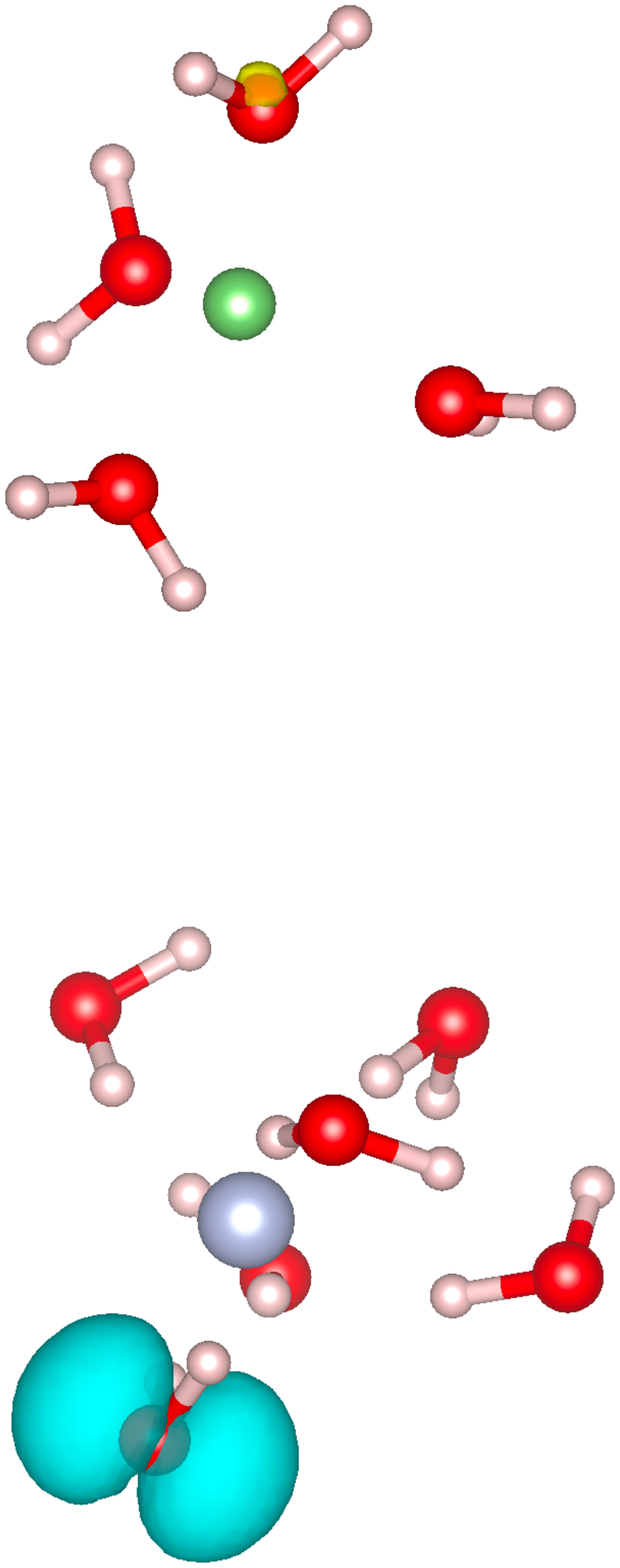}
        \caption{CIS}
        \label{fig:cis}
    \end{subfigure}
    \qquad
    \begin{subfigure}{0.3\textwidth}
        \centering
        \includegraphics[scale=0.295]{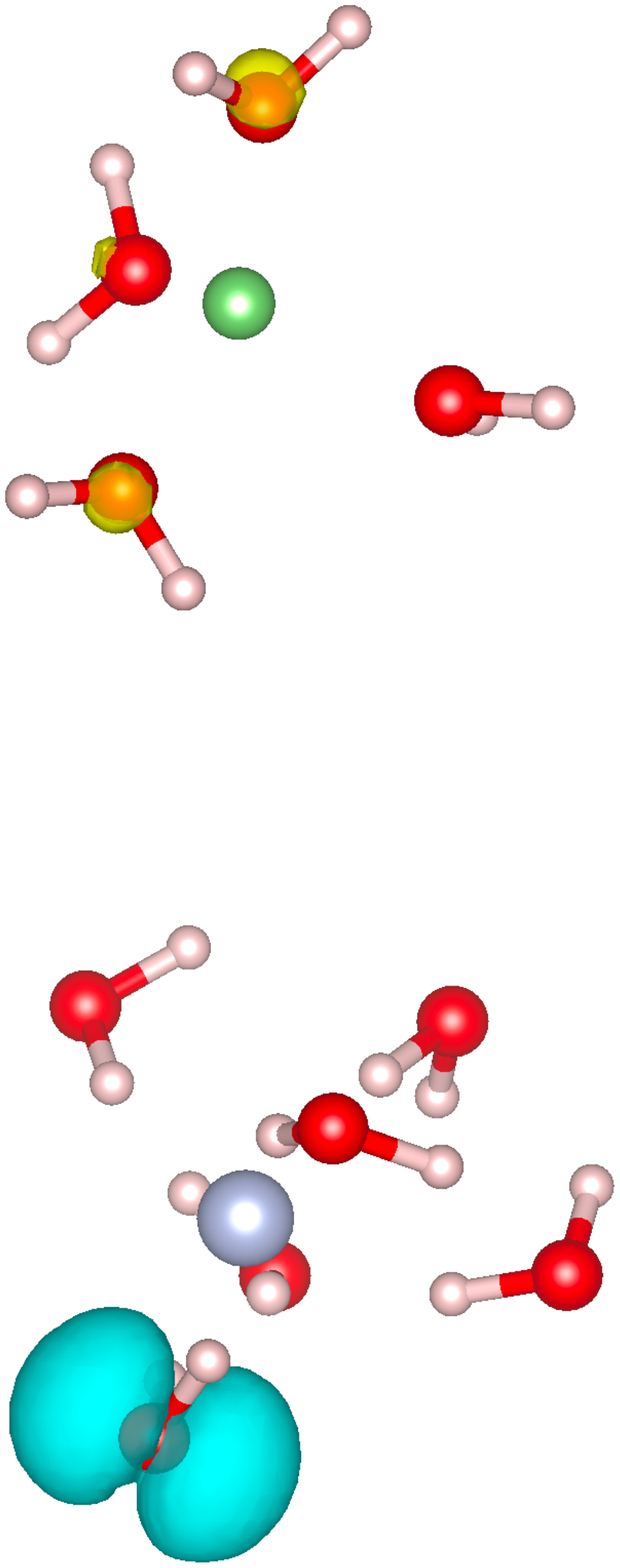}
        \caption{TDDFT}
        \label{fig:wb97x}
    \end{subfigure}
    \qquad
    \begin{subfigure}{0.3\textwidth}
        \centering
        \includegraphics[scale=0.30]{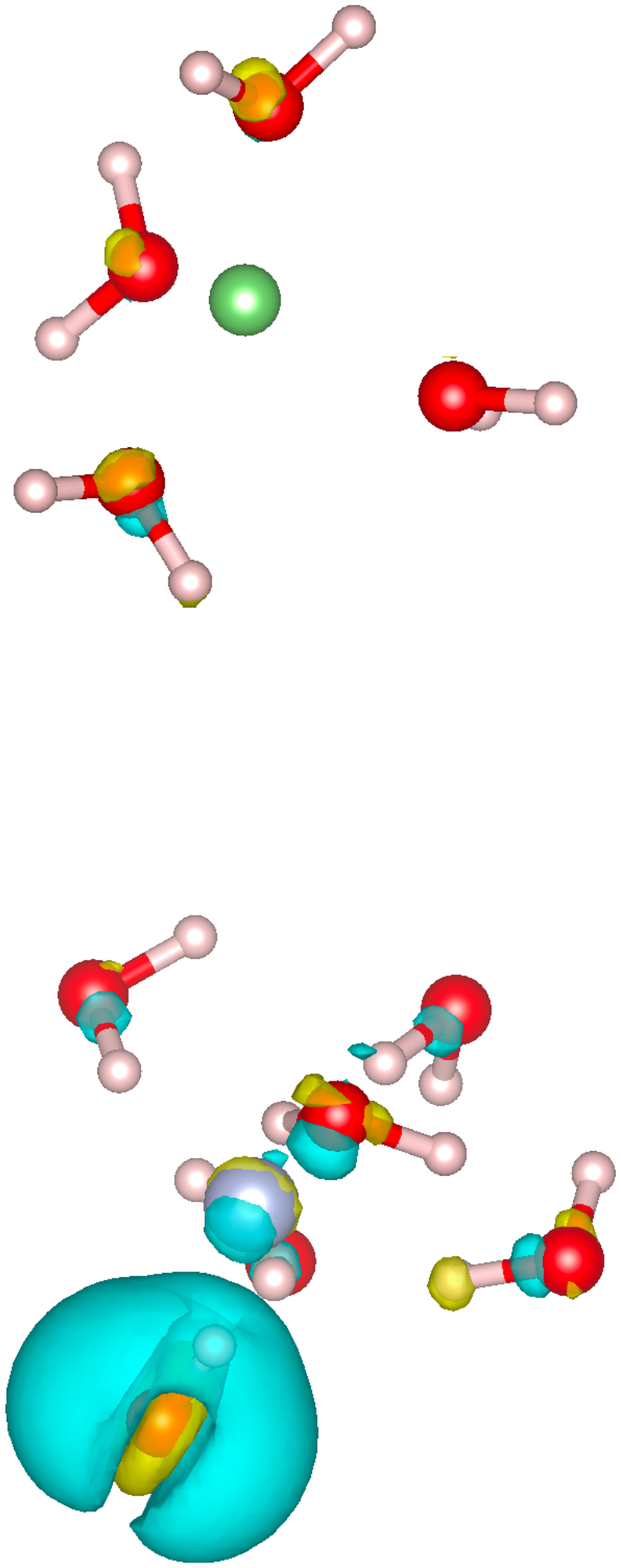}
        \caption{ESMF}
        \label{fig:esmf}
    \end{subfigure}
    \caption{Isosurface plots for the charge density changes following
             the charge transfer excitation
             that moves an electron from the lower F cluster to
             the upper Li cluster, with blue
             surfaces showing charge depletion relative to
             the ground state and yellow surfaces
             showing charge accumulation.
             We compare the results using CIS, TDDFT, and ESMF 
             in Figure \ref{fig:cis}, \ref{fig:wb97x}, and \ref{fig:esmf}. 
             TDDFT employed the $\omega$B97X functional.
             Note that the primary orbital on the Li cluster that accepts
             the transferred electron is very diffuse, and so at the isosurface
             value that makes charge-relaxation effects visible in the lower
             cluster the charge accumulation on the Li cluster is not visible.
             See the Supplemental Materials for a plot with an isosurface value that makes
             the Li cluster charge accumulation more clear.
            }
    \label{fig:diff_den_comp}
\end{figure*}

\begin{figure*}
    \centering
    \begin{subfigure}{0.3\textwidth}
        \centering
        \includegraphics[scale=0.25]{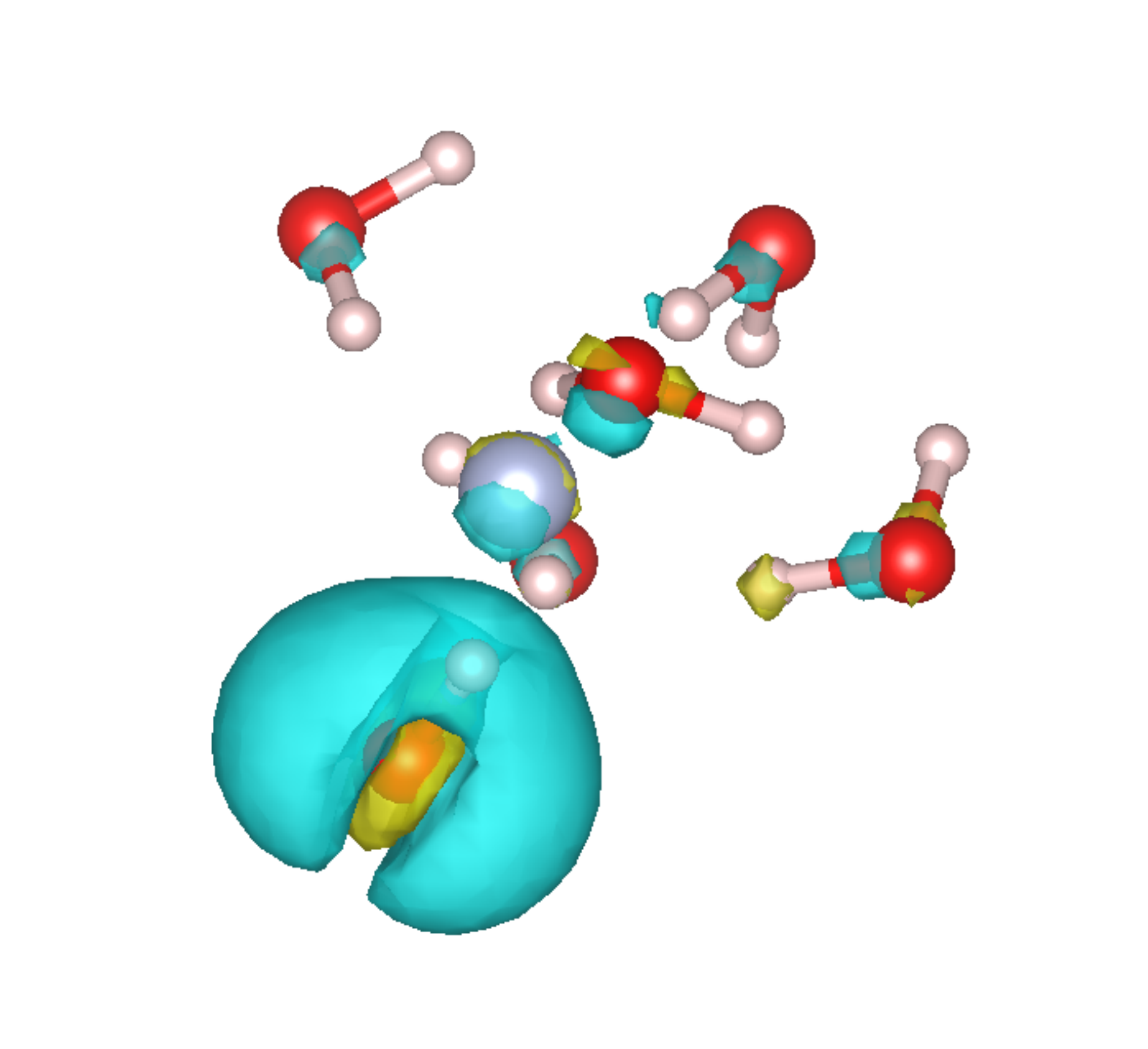}
        \caption{ROHF}
        \label{fig:rohf}
    \end{subfigure}
    \qquad
    \begin{subfigure}{0.3\textwidth}
        \centering
        \includegraphics[scale=0.24]{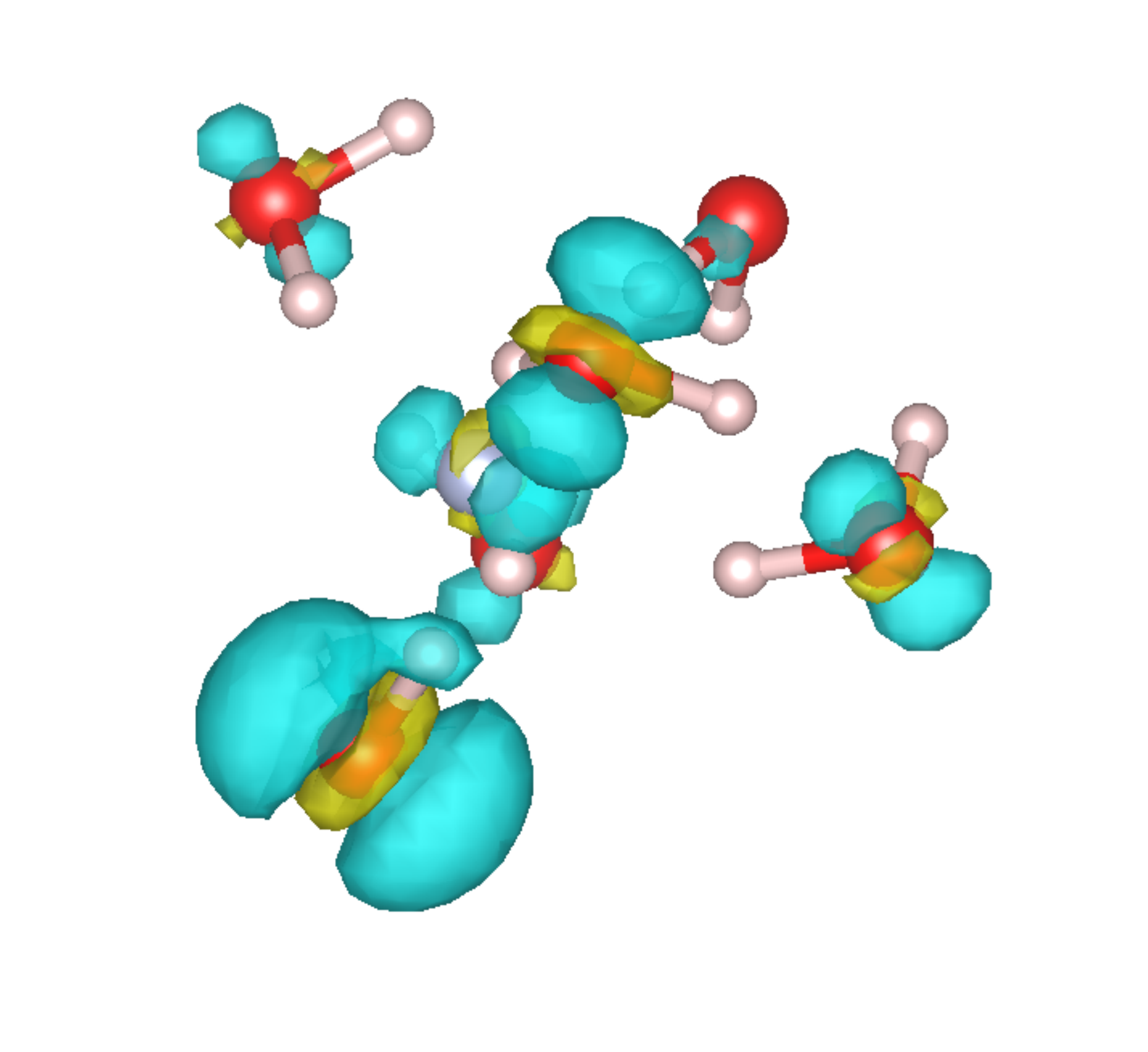}
        \caption{UKS}
        \label{fig:uks}
    \end{subfigure}
    \qquad
    \begin{subfigure}{0.3\textwidth}
        \centering
        \includegraphics[scale=0.23]{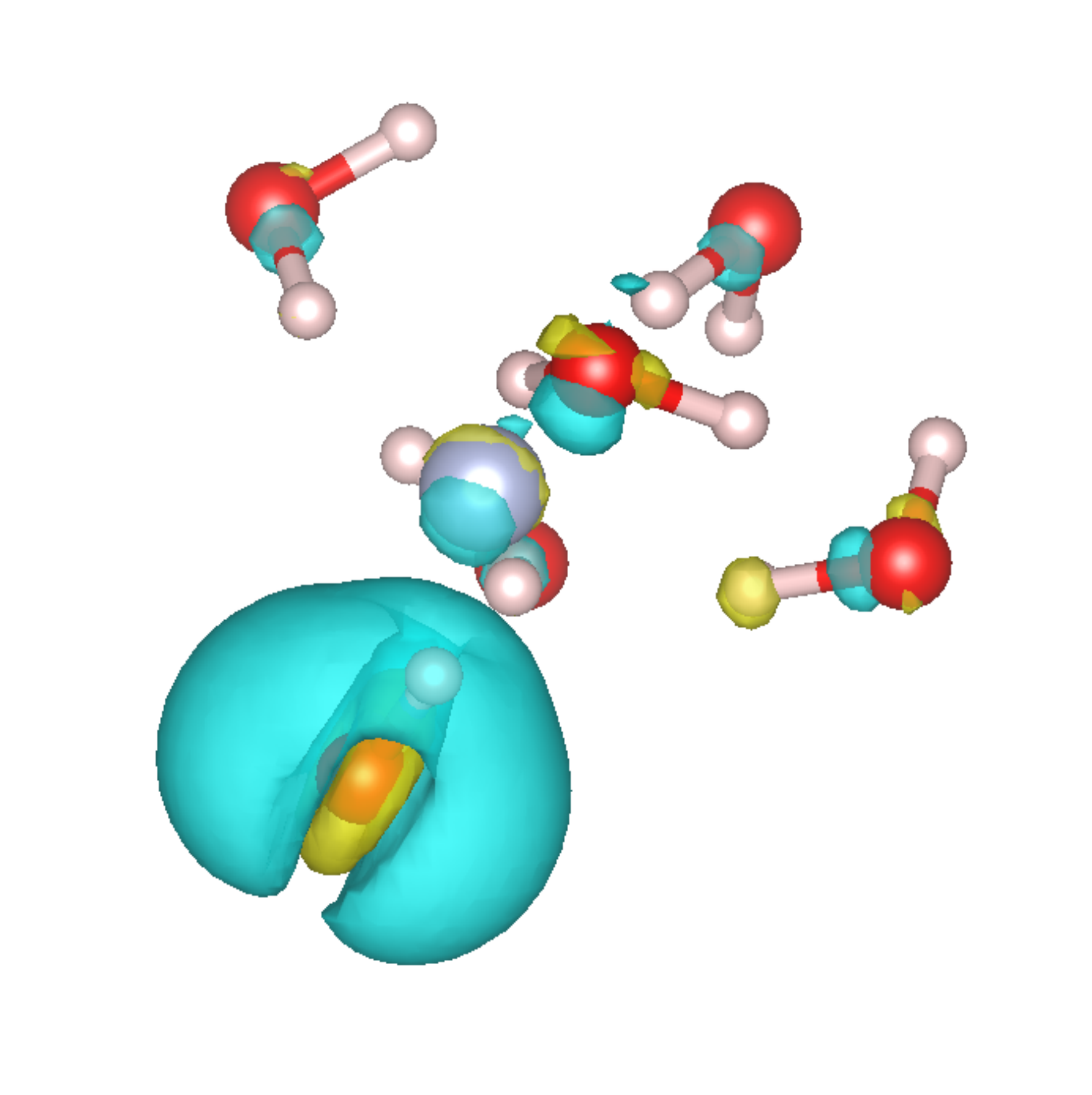}
        \caption{ESMF}
        \label{fig:esmf_lower}
    \end{subfigure}
     \caption{
       As for Figure \ref{fig:diff_den_comp}, but showing ROHF and UKS ground-state
       charge density differences between the neutral and negatively charged F cluster
       when modeled in isolation alongside ESMF's charge density differences
       when it treats the charge transfer state in the full two-cluster system.
       We compare the results using ROHF, UKS, and ESMF 
       in Figure \ref{fig:rohf}, \ref{fig:uks}, and \ref{fig:esmf_lower}. 
       UKS employed the $\omega$B97X functional.
     }
    \label{fig:diff_den_comp_lower}
\end{figure*}

Finally, in order to verify that the gain we obtained in efficiency does not come with a
loss in accuracy from the shell-pair screening, 
we show the predicted excitation energies of Cl$^-$-H$_2$O
and NH$_3$-F$_2$ in Table \ref{tab:ee_compare}.
We found that the predictions match to better than 10$^{-4}$eV
between the two implementations, indicating that there was no meaningful
accuracy loss at the screening threshold we employed.

\subsection{Minimally-solvated charge transfer}

\begin{table*}
\caption{
  Differences in atomic Mulliken charges for atoms in
  the lower cluster between the
  excited (lower cluster neutral) and ground
  (lower cluster negatively charged) states.
  For UKS, ROHF, and IP/EA-EOM-CCSD, we report the results for the
  corresponding charge states of the lower cluster in isolation
  (see text).
  DFT methods used the $\omega$B97X functional.
  \label{tab:pop_analysis}
}
\begin{tabular}{c c c c c c c c c c c c c c c c c c c c}
\hline\hline
Method & F & O & H & H & O & H & H & O & H & H & O & H & H & O & H & H & O & H & H \\
\hline
CIS &   0.01 &  0.01 & 0.00 & 0.00 & 0.00 & 0.00 & 0.00 & 0.00 & 0.00 & 0.00 &  0.96 &  0.01 &  0.01 & 0.00 & 0.00 & 0.00 & 0.00 & 0.00 & 0.00 \\
TDDFT &   0.01 & 0.00 & 0.00 & 0.00 & 0.00 & 0.00 & 0.00 & 0.00 & 0.00 & 0.00 &  0.96 &  0.01 &  0.01 & 0.00 & 0.00 & 0.00 & 0.00 & 0.00 & 0.00 \\
UKS &   0.06 &  0.04 & 0.00 &  0.02 &  0.11 &  0.01 &  0.03 &  0.10 &  0.03 &-0.01 &  0.38 &  0.06 &  0.05 &  0.03 &-0.01 &  0.01 &  0.08 &  0.01 &  0.03 \\
ROKS &   0.03 &  0.14 & 0.02 &  0.03 &  0.09 &  0.00 &  0.02 &  0.07 &  0.03 &-0.01 &  0.32 &  0.06 &  0.05 &  0.04 &-0.01 &  0.01 &  0.06 &  0.00 &  0.02 \\
ROHF &   0.01 &  0.02 &-0.01 &  0.02 &  0.02 &-0.02 &  0.01 &  0.02 &  0.03 &-0.03 &  0.72 &  0.10 &  0.09 &  0.02 &-0.01 &  0.01 &  0.02 &-0.02 &  0.01 \\
ESMF &   0.01 &  0.03 &-0.01 &  0.02 &  0.02 &-0.02 &  0.01 &  0.01 &  0.03 &-0.04 &  0.71 &  0.11 &  0.09 &  0.03 &-0.01 & 0.00 &  0.02 &-0.02 &  0.01 \\
CCSD &  0.02 &  0.01 &-0.01 &  0.01 &  0.01 &-0.01 & 0.00 &  0.01 &  0.01 &-0.02 &  0.78 &  0.09 &  0.08 &  0.01 &-0.01 & 0.00 &  0.01 &-0.01 & 0.00 \\
\hline\hline
\vspace{2mm}
\end{tabular}
\end{table*}

With these efficiency improvements in hand, we are able to make an initial investigation
of ESMF's predictions for how charge transfers in a minimally solvated environment.
To this purpose, we study the system shown in Figure \ref{fig:lif_geo},
in which a Li atom and a F atom are each surrounded by a small collection
of water molecules.
Although the geometry optimizations used a different basis (see figure), all subsequent
calculations were performed in the cc-pVDZ orbital basis.
When the two clusters are simulated as a single system with ground state methods,
the F and Li clusters carry negative and positive net charges, respectively.
Our purpose here is to investigate the predictions that various excited state methods
make for how the charge density changes in the (lowest-energy singlet) charge
transfer excitation that returns both clusters to charge neutrality.


As seen in Figure \ref{fig:diff_den_comp}, the charge density changes
predicted by ESMF differ noticeably from those of CIS and
$\omega$B97X-based \cite{chai2008systematic} TDDFT.
Although all three methods (and all others tested, see below) agree that
the transferred electron comes from a lone pair on the lower-left water molecule,
ESMF also displays two types of significant orbital-relaxation effects that
CIS and TDDFT fail to capture.
First, electron density in the OH bonds on the affected water molecule is seen
to shift towards the oxygen atom, showing a net charge depletion on the hydrogen
atoms and some net charge accumulation close to the oxygen atom's center.
Second, the other water molecules show a polarization of their electron densities
in response to the newly created hole on the lower-left water.
As seen in Table \ref{tab:pop_analysis}, these relaxation effects create
changes in the Mulliken populations for ESMF that differ significantly from
those shown by CIS and TDDFT, both of which
predict only significant changes to the lower-left oxygen atom.

\begin{table}[b]
\caption{Comparison of excitation energies (eV) for the Li-F system's
         charge transfer excitation.
         See text for details of the the CCSD approach.
\label{tab:ee}
}
\begin{tabular}{c c}
\hline\hline
Method & Excitation Energy \\
\hline
CIS & 8.82  \\
TDDFT/$\omega$B97X & 6.12 \\
ESMF & 5.92 \\
CCSD & 6.64 \\
\hline\hline
\vspace{2mm}
\end{tabular}
\end{table}

At this point, we take advantage of the fact that the two clusters are well
separated in order to perform IP/EA-EOM-CCSD\cite{Mukherjee1988,Gauss1994,Barlett1995} on the separate clusters in isolation
in order to produce a high-level benchmark on what types of charge-relaxation
effects other methods should display.
As seen in Table \ref{tab:pop_analysis}, the coupled cluster predictions for
the Mulliken population changes on both the lower-left water and the other
water molecules agree reasonably well with the ESMF predictions while
disagreeing with the CIS and TDDFT predictions.
In particular, CIS and TDDFT fail to predict the significant relaxation near
the lower-left water's hydrogen atoms as well as the degree to which the charge
densities on the other water molecules shift in response to the newly created hole.
To verify that these errors are due to the additional approximations
(namely a lack of secondary orbital relaxation) associated with CIS and TDDFT
as excited state linear-response methods, we have also performed ground state
restricted open-shell HF (ROHF) and $\omega$B97X-based unrestricted Kohn-Sham (UKS)
calculations on the anionic and neutral lower cluster in isolation.
As seen in Figure \ref{fig:diff_den_comp_lower} and Table \ref{tab:pop_analysis},
these methods both display significant relaxation effects upon creation of the hole
on the lower-left water, as should be expected of self-consistent field methods upon
the removal of an electron.
However, the ROHF and UKS predictions are quite different from each other,
with the latter incorrectly delocalizing the hole across all of the water
molecules in the lower cluster in a clear display of the difficulties posed
by DFT's self-interaction error, which is known to
over-stabilize delocalized states. \cite{Siegbahn2005,Yang2008}
ROHF, ESMF, and coupled cluster, in contrast, are by construction free of
self-interaction errors and keep the hole localized on the lower-left water
molecule, with the other water molecules showing much smaller net population
changes while still correctly shifting their electron clouds to re-polarize
after the creation of the hole.
Note especially that the ROHF and ESMF results are in close agreement,
which is not surprising given that the spin-purity offered by ESMF has only a
very small energetic effect in this type of long-range charge transfer excitation.
The same logic explains why the ROKS results are very similar to those of UKS,
displaying the same clear signs of self-interaction-induced over-delocalization.

As a final note in this system, we remind the reader that ESMF, like HF, should not be
expected to deliver highly accurate energetics as it neglects weak correlation effects.
For the cluster under study, excitation energy estimates for different methods are
shown in Table \ref{tab:ee}.
Note that due to the large system size, we did not pursue a direct
EOM-CCSD calculation in the full system.
Instead, we created an estimate of the coupled
cluster excitation energy via the expression
\begin{equation}
    \label{eqn:ccsd_es}
    \Delta E=\mathrm{[EA]_{upper}}-\mathrm{[IP]_{lower}}+1/R
\end{equation}
in which $\mathrm{[EA]_{upper}}$ is the electron affinity of the upper cluster in
isolation and $\mathrm{[IP]_{lower}}$ is the ionization potential of the lower cluster
in isolation, as predicted by IP/EA-EOM-CCSD.
The $1/R$ term is used to estimate the Coulomb attraction of the two charged clusters
in ground state, with $R$ computed as the distance between Li and F.
As we have seen in many other systems, \cite{Shea2019}
ESMF appears to underestimate the excitation energy.
To reach quantitative energetics, a correlation correction such as the one
provided by ESMP2 \cite{Shea2019} is clearly in order.
While the current formulation and implementation of that theory
is too expensive to use in a system of this size, work exploring more affordable
formulations is ongoing.

\subsection{More Charge Density Tests}
\label{sec:density_changes}
\begin{table}[t]
\caption{
  Differences in atomic Mulliken populations for atoms between the
  excited and ground states of NaCl.
  DFT methods used the $\omega$B97X functional.
  \label{tab:pop_analysis_nacl}
}
\begin{tabular}{c c c c}
\hline\hline
Method & Na & Cl & total error  \\
\hline
CIS &     -0.93 & 0.93 & 0.41\\
TDDFT &   -0.93 & 0.93 & 0.39\\
ROKS &    -0.57 & 0.57 & 0.31\\
ESMF &    -0.69 & 0.69 & 0.08\\
EOM-CCSD &    -0.73 & 0.73 & N/A \\
\hline\hline
\vspace{2mm}
\end{tabular}
\end{table}

In light of ESMF's success in predicting charge density changes in
the previous section, we now turn our attention to how accurately
the method predicts charge density changes in four
additional charge transfer excitations.
We examine charge transfer excitations in
the NaCl molecule, the
Cl$^-$-(H$_2$O)$_3$ system,
the C$_2$H$_4$-C$_2$F$_4$ dimer,
and the aminophenol molecule. 
In these excitations,
the charge flow is from Cl to Na, from Cl$^-$ to H$_2$O, from C$_2$H$_4$ to 
C$_2$F$_4$, and from aminophenol's OH and NH$_2$ groups to its benzene ring.
Note that we provide the molecular structures and 
Cartesian coordinates of these test systems in the Supplemental Materials. 

Tables \ref{tab:pop_analysis_nacl}, \ref{tab:pop_analysis_cl_3h2o},
\ref{tab:pop_analysis_c2h4_c2f4}, and \ref{tab:pop_analysis_aminophenol}
show how well ESMF, TDDFT, CIS, ROKS, and CIS do at predicting charge density
differences in comparison to EOM-CCSD.
Note that we use standard EOM-CCSD here on the full systems, as they
are small enough for this to be practical,
and that in some cases we use $\Delta$SCF instead of ROKS due to
QChem's current limitation of ROKS to HOMO/LUMO excitations.
As their charge densities should be the same, we use the ROKS
and $\Delta$SCF labels interchangeably in this section.
We find that ESMF is consistently more accurate in these predictions
than either CIS or TDDFT.
When comparing ESMF and ROKS, we find that in three systems
---  Cl$^-$-(H$_2$O)$_3$,
     C$_2$H$_4$-C$_2$F$_4$,
     and aminophenol ---
these methods' accuracies are similar.
However, for the charge density changes in NaCl and in
the solvated charge transfer example of the previous section,
ESMF is substantially more accurate than ROKS.
The poorer performance of CIS and TDDFT can largely be explained by
their inability to relax the orbitals not directly involved in the excitation.
ROKS, on the other hand, does contain these types of relaxations
through its $\Delta$SCF approach, but in some cases self-interaction error
causes it to significantly overestimate the delocalization of the hole.
We emphasize that we have employed a range-separated hybrid functional
($\omega$B97X) for TDDFT and ROKS in order to maximize their ability
to deal successfully with charge transfer.
Nonetheless, we find that ESMF is more reliable for predicting
how the charge density changes in these charge transfer excitations.

\section{Conclusion}
\label{sec:conclusion}

We have presented explicit analytic expressions for the derivatives required
for ESMF wave function optimization.
Although these expressions are somewhat long winded, the upshot is that the
necessary derivatives can be evaluated through nine Fock matrix builds and
a collection of (much less expensive) single-particle matrix operations.
This formulation allows ESMF to immediately benefit from methods that improve
the efficiency of Fock builds, a situation we have exploited via the
shell-pair screening strategy in order to construct a thread-parallel ESMF
implementation that is multiple orders of magnitude faster than the previous
implementation, which relied on automatic differentiation.
We then applied our method to study a Li-F charge transfer system
in which these atoms were surrounded by minimal water solvation shells.
By comparing against IP/EA-EOM-CCSD, we found that ESMF's predictions for
how the charge transfer excitation changes the electron cloud
were qualitatively more accurate than the predictions of CIS, TDDFT, UKS, or ROKS.
In particular, ESMF correctly captured post-excitation orbital relaxation effects
and, by virtue of being free of self-interaction errors, did not erroneously
delocalize the hole.

Looking forward, these findings point to multiple priorities for further research.
First, by giving ESMF access to larger systems, they make the development
of an $N^5$-scaling post-ESMF perturbation theory (as opposed to the $N^7$-scaling
of the existing \cite{Shea2018} version) even more pressing.
Happily, a path forward in this direction has been found and will be published soon.
\cite{n5esmp2}
Second, we would expect the recently-introduced density functional extension
to ESMF to show similar efficiency improvements upon development of Fock-build-based
analytic gradients.
Finally, in light of the efficiency offered by the successful generalization
of the geometric direct minimization method to excited state orbital optimization,
\cite{hait2019delSquared}
it seems likely that employing our new analytic gradients in a custom-tailored
quasi-Newton method would lead to further efficiency gains.

\section{Supplemental Material}

See supplementary material for the molecular geometries, Cartesian coordinates, and overlap 
integrals between ground and excited states. 

\begin{table*}[h]
\caption{
  Differences in atomic Mulliken populations for atoms between the
  excited and ground states of Cl$^-$-(H$_2$O)$_3$.
  DFT methods used the $\omega$B97X functional.
  \label{tab:pop_analysis_cl_3h2o}
}
\begin{tabular}{c c c c c c c c c c c c}
\hline\hline
Method & Cl & O & H & H & O & H & H & O & H & H & total error \\
\hline
CIS &    0.86 & 0.04 & -0.21 & -0.09 & 0.07 & -0.11 & -0.24 & 0.04 & -0.26 & -0.09 & 0.32 \\
TDDFT &  0.88 & 0.01 & -0.18 & -0.13 & 0.04 & -0.14 & -0.19 & 0.02 & -0.19 & -0.13 & 0.25\\
ROKS &   0.78 & 0.07 & -0.21 & -0.12 & 0.08 & -0.12 & -0.19 & 0.09 & -0.24 & -0.13 & 0.17 \\
ESMF &   0.85 & 0.06 & -0.19 & -0.14 & 0.07 & -0.15 & -0.21 & 0.08 & -0.24 & -0.15 & 0.16 \\
EOM-CCSD &   0.80 & 0.06 & -0.18 & -0.13 & 0.08 & -0.14 & -0.20 & 0.06 & -0.21 & -0.13 & N/A \\
\hline\hline
\vspace{2mm}
\end{tabular}
\end{table*}

\begin{table*}[h!]
\caption{
  Differences in atomic Mulliken populations for atoms between the
  excited and ground states of C$_2$H$_4$-C$_2$F$_4$.
  DFT methods used the $\omega$B97X functional.
  \label{tab:pop_analysis_c2h4_c2f4}
}
\begin{tabular}{c c c c c c c c c c c c c c}
\hline\hline
Method & C & C & H & H & H & H & C & C & F & F & F & F & total error \\
\hline
CIS &    0.49 & 0.49 &  0.00 & 0.00 &  0.00 &  0.00 & -0.46 & -0.46 & -0.02 & -0.02 & -0.02 & -0.02 & 1.30\\
TDDFT &  0.49 & 0.49 &  0.00 & 0.00 &  0.00 &  0.00 & -0.44 & -0.44 & -0.03 & -0.03 & -0.03 & -0.03 & 1.19\\
ROKS &   0.22 & 0.22 &  0.16 & 0.13 &  0.16 &  0.13 & -0.28 & -0.28 & -0.10 & -0.12 & -0.10 & -0.12 & 0.52 \\
ESMF &   0.24 & 0.24 &  0.15 & 0.12 &  0.15 &  0.12 & -0.28 & -0.28 & -0.10 & -0.12 & -0.10 & -0.12 & 0.45\\
EOM-CCSD &   0.30 & 0.30 &  0.11 & 0.09 &  0.11 & 0.09 & -0.33 & -0.33 & -0.08 & -0.09 & -0.08 & -0.09 & N/A\\
\hline\hline
\vspace{2mm}
\end{tabular}
\end{table*}

\begin{table*}[h!]
\caption{
  Differences in atomic Mulliken populations for atoms between the
  excited and ground states of aminophenol.
  DFT methods used the $\omega$B97X functional.
  \label{tab:pop_analysis_aminophenol}
}
\begin{tabular}{c c c c c c c c c c c c c c c c c}
\hline\hline
Method & H & C & C & H & C & O & C & H & C & H & C & N & H & H & H & total error  \\
\hline
CIS &     -0.00 &  -0.10 &  -0.14 & -0.00 & 0.10 & 0.05 & -0.13 &  0.00 & -0.07 & 0.00 & 0.12 & 0.12 & 0.00 & 0.00 & 0.00 & 0.57 \\
TDDFT &   -0.00 &  -0.11 &  -0.14 & -0.00 & 0.11 & 0.09 & -0.13 &  0.00 & -0.09 & 0.00 & 0.07 & 0.19 & 0.00 & 0.00 & 0.00 & 0.60 \\
ROKS &    -0.02 &  -0.04 &  -0.06 & -0.02 & 0.05 & 0.04 & -0.06 & -0.02 & -0.03 & -0.02 &  0.00 & 0.11 & 0.01 & 0.03 & 0.03 & 0.13 \\
ESMF &    -0.01 &  -0.04 &  -0.07 & -0.02 & 0.07 & 0.03 & -0.07 & -0.02 & -0.03 & -0.01 &  0.00 & 0.11 & 0.01 & 0.03 & 0.03 & 0.18 \\
EOM-CCSD &    -0.01 &  -0.05 &  -0.06 & -0.01 & 0.03 & 0.05 & -0.05 & -0.01 & -0.04 & -0.01 &  0.00 & 0.12 & 0.01 & 0.02 & 0.02 & N/A \\
\hline\hline
\vspace{2mm}
\end{tabular}
\end{table*}


\begin{acknowledgments}
This work was supported by the National Science Foundation's
CAREER program under Award Number 1848012.
L.Z.\ acknowledges additional support from the Dalton Fellowship
at the University of Washington. 
The Berkeley Research Computing Savio cluster performed the calculations.
\end{acknowledgments}

\bibliography{FF_ESMF}

\clearpage

\begin{appendices}
We provide the analytic expression of ESMF energy and target function derivatives in 
this section. Although the final results look tedious, the derivation is trivial as 
the ESMF energy and target function only involve matrix trace and products. Here 
we take the derivative of the first term in $E_1$ and $E_2$ with respect to 
$\Theta_{kp}$ for example,
\begin{equation}
    \label{eqn:eng1_der_theta_kp}
    \begin{split}
        &\frac{\partial}{\partial\Theta_{kp}}\left(2N_2\mathrm{Tr}[\bm{\Theta}\textbf{G}\bm{\Theta}^T]\right) \\
        &=\frac{\partial}{\partial\Theta_{kp}}\left(2N_2\sum_i{[\bm{\Theta}\textbf{G}\bm{\Theta}]_{ii}}\right) \\
        &=\frac{\partial}{\partial\Theta_{kp}}\left(2N_2\sum_i{\sum_{rs}{\Theta_{ir}G_{rs}\Theta_{is}}}\right) \\
        &=2N_2\sum_i{\sum_{rs}{\Theta_{ir}G_{rs}\delta_{ki}\delta_{ps}}}+2N_2\sum_i{\sum_{rs}{\delta_{ki}\delta_{rp}G_{rs}\Theta_{is}}} \\
        &=2N_2\sum_{r}{\Theta_{kr}G_{rp}}+2N_2\sum_{r}{G_{ps}\Theta_{ks}} \\
        &=4N_2\left(\bm{\Theta}\textbf{G}\right)_{kp} \\
    \end{split}
\end{equation}

\begin{equation}
    \label{eqn:eng2_der_theta_kp}
    \begin{split}
        &\frac{\partial}{\partial\Theta_{kp}}\left(N_2\textbf{F}[\bm{\Theta}^T\bm{\Theta}]\cdot\left(\bm{\Theta}^T\bm{\Theta}\right)\right) \\
        &=\frac{\partial}{\partial\Theta_{kp}}\left(N_2\sum_{rs}{\textbf{F}[\bm{\Theta}^T\bm{\Theta}]_{rs}\left(\bm{\Theta}^T\bm{\Theta}\right)_{rs}}\right) \\
        &=N_2\sum_{rs}{\frac{\textbf{F}[\bm{\Theta}^T\bm{\Theta}]_{rs}}{\Theta_{kp}}\left(\bm{\Theta}^T\bm{\Theta}\right)_{rs}}+N_2\sum_{rs}{\textbf{F}[\bm{\Theta}^T\bm{\Theta}]_{rs}\frac{\left(\bm{\Theta}^T\bm{\Theta}\right)_{rs}}{\Theta_{kp}}}
    \end{split}
\end{equation}
The derivative of the Fock matrix is
\begin{equation}
    \label{eqn:F_der}
    \begin{split}
        &\frac{\partial\textbf{F}[\bm{\Theta}^T\bm{\Theta}]_{rs}}{\partial\Theta_{kp}} \\
        &=\frac{\partial}{\partial\Theta_{kp}}\left(2\textbf{J}[\bm{\Theta}^T\bm{\Theta}]_{rs}-\textbf{K}[\bm{\Theta}^T\bm{\Theta}]_{rs}\right) \\
        &=2\sum_{v}{\Theta_{kv}(pv|rs)}+2\sum_{u}{\Theta_{ku}(pu|rs)} \\
        &-\sum_{v}{\Theta_{kv}(rp|sv)}-\sum_{u}{\Theta_{ku}(ru|sp)} \\
    \end{split}
\end{equation}

Plugging this into the first term of Equation \ref{eqn:eng1_der_theta_kp}, we obtained
\begin{equation*}
    \label{eqn:eng2_der_theta_kp2}
    \begin{split}
        &N_2\sum_{rs}{\frac{\textbf{F}[\bm{\Theta}^T\bm{\Theta}]_{rs}}{\Theta_{kp}}\left(\bm{\Theta}^T\bm{\Theta}\right)_{rs}} \\
        &=N_2\sum_{rs}{\left(\bm{\Theta}^T\bm{\Theta}\right)_{rs}[2\sum_{v}{\Theta_{kv}(pv|rs)}+2\sum_{u}{\Theta_{ku}(pu|rs)}]} \\
        &-N_2\sum_{rs}{\left(\bm{\Theta}^T\bm{\Theta}\right)_{rs}[\sum_{v}{\Theta_{kv}(rp|sv)}+\sum_{u}{\Theta_{ku}(ru|sp)}]} \\
        &=N_2\sum_{v}{\Theta_{kv}\textbf{F}[\bm{\Theta}^T\bm{\Theta}]_{pv}}+N_2\sum_{u}{\Theta_{ku}\textbf{F}[\bm{\Theta}^T\bm{\Theta}]_{pu}} \\
        &=2N_2\left(\bm{\Theta}\textbf{F}[\bm{\Theta}^T\bm{\Theta}]\right)_{kp}
    \end{split}
\end{equation*}

The second term of Equation \ref{eqn:eng1_der_theta_kp} is
\begin{equation}
    \label{eqn:eng2_der_theta_kp3}
    \begin{split}
        &N_2\sum_{rs}{\textbf{F}[\bm{\Theta}^T\bm{\Theta}]_{rs}\frac{\left(\bm{\Theta}^T\bm{\Theta}\right)_{rs}}{\Theta_{kp}}} \\
        &\sum_{rs}{\textbf{F}[\bm{\Theta}^T\bm{\Theta}]_{rs}\sum_i{\delta_{ki}\delta_{pr}\Theta_{is}}}+\sum_{rs}{\textbf{F}[\bm{\Theta}^T\bm{\Theta}]_{rs}\sum_i{\Theta_{ir}\delta_{ki}\delta_{ps}}} \\
        &=2N_2\left(\bm{\Theta}\textbf{F}[\bm{\Theta}^T\bm{\Theta}]\right)_{kp} \\
    \end{split}
\end{equation}

Putting these together, we obtain
\begin{equation}
    \label{eqn:eng2_der_theta_kp4}
    \frac{\partial}{\partial\Theta_{kp}}\left(N_2\textbf{F}[\bm{\Theta}^T\bm{\Theta}]\cdot\left(\bm{\Theta}^T\bm{\Theta}\right)\right)=4N_2\left(\bm{\Theta}\textbf{F}[\bm{\Theta}^T\bm{\Theta}]\right)_{kp}
\end{equation}
The other terms of the energy and objective function derivatives could be computed in similar 
ways, and their expression are as follows.

The derivative 
with respect to $c_0$ is,
\begin{equation}
    \label{eqn:eng_der_c0}
    \begin{split}
    \frac{\partial E}{\partial c_0}&=\left(4c_0\mathrm{Tr}[\bm{\Theta}\textbf{ G}\bm{\Theta}^T]+4\mathrm{Tr}[\bm{\Theta}\textbf{G}\bm{\Gamma}^T\bm{\sigma}^T]\right. \\
    &\left.+2c_0\textbf{F}[\bm{\Theta}^T\bm{\Theta}]\cdot\left(\bm{\Theta}^T\bm{\Theta}\right)+4\textbf{F}[\bm{\Theta}^T\bm{\Theta}]\cdot\left(\bm{\Theta}^T\bm{\sigma\Gamma}\right)\right)/N_2 \\
    &-2c_0E/N_2 \\
    \end{split}
\end{equation}

The derivative with respect to $\sigma_{jb}$ is,
\begin{equation}
    \label{eqn:eng_der_sigma_jb}
    \begin{split}
        \frac{\partial E}{\partial \sigma_{jb}}&=\left(4c_0\left(\bm{\Theta} \textbf{G}\bm{\Gamma}^T\right)_{jb}+4\left(\bm{\sigma\Gamma}\textbf{G}\bm{\Gamma}^T\right)_{jb}\right. \\
        &-4\left(\bm{\Theta}\textbf{G}\bm{\Theta}^T\bm{\sigma}\right)_{jb}+8\mathrm{Tr}[\bm{\Theta} \textbf{G}\bm{\Theta}^T]\sigma_{jb} \\
        &+4c_0\left(\bm{\Theta} \textbf{F}[\bm{\Theta}^T\bm{\Theta}]\bm{\Gamma}^T\right)_{jb}+4\left(\bm{\sigma \Gamma} \textbf{F}[\bm{\Theta}^T\bm{\Theta}]\bm{\Gamma}^T\right)_{jb} \\
        &-4\left(\bm{\Theta} \textbf{F}[\bm{\Theta}^T\bm{\Theta}]\bm{\Theta}^T\bm{\sigma}\right)_{jb}+4\left(\bm{\Theta} \textbf{F}[\bm{\Theta}^T\bm{\sigma \Gamma}]\bm{\Gamma}^T\right)_{jb} \\
        &\left.+4\sum_{pq}{\left(\textbf{F}[\bm{\Theta}^T\bm{\Theta}]\bm{\Theta}^T\bm{\Theta}\right)_{pq}}\sigma_{jb}\right)/N_2 \\
        &-4E\sigma_{jb}/N_2 \\
    \end{split}
\end{equation}

The derivative with respect to $\Theta_{kp}$ is,
\begin{equation}
    \label{eqn:eng_der_ukp}
    \begin{split}
        \frac{\partial E}{\partial \Theta_{kp}}&=\left(4N_2\left(\bm{\Theta}\textbf{G}\right)_{kp}+4c_0\left(\bm{\sigma \Gamma} \textbf{G}\right)_{kp}\right.\\
        &-4\left(\bm{\sigma\sigma}^T\bm{\Theta}\textbf{G}\right)_{kp}+4N_2\left(\bm{\Theta} \textbf{F}[\bm{\Theta}^T\bm{\Theta}]\right)_{kp} \\
        &+4c_0\left(\bm{\sigma\Gamma}\textbf{F}[\bm{\Theta}^T\bm{\sigma\Gamma}]\right)_{kp} \\
        &+4c_0\left(\bm{\Theta}\left(\textbf{F}[\bm{\Theta}^T\bm{\sigma\Gamma}]-F[\bm{\Theta}^T\bm{\sigma\Gamma}]^T\right)\right)_{kp} \\
        &+2\left(\bm{\Theta}\left(\textbf{F}[\textbf{A}]-\textbf{F}[\textbf{A}]^T\right)\right)_{kp}-4\left(\bm{\sigma\sigma}^T\bm{\Theta}\textbf{F}[\bm{\Theta}^T\bm{\Theta}]\right)_{kp} \\
        &\left.+4\left(\bm{\sigma\Gamma}\textbf{F}[\bm{\Theta}^T\bm{\sigma\Gamma}]\right)_{kp}/N_2\right) \\
    \end{split}
\end{equation}

The derivative with respect to $\Gamma_{cp}$ is,
\begin{equation}
    \label{eqn:eng_der_ucp}
    \begin{split}
        \frac{\partial E}{\partial\Gamma_{cp}}&=\left(4c_0\left(\bm{\sigma}^T\bm{\Theta} \textbf{G}\right)_{cp}+4\left(\bm{\sigma}^T\bm{\sigma\Gamma}\textbf{G}\right)_{cp}\right. \\
        &+4c_0\left(\bm{\sigma}^T\bm{\Theta}\textbf{F}[\bm{\Theta}^T\bm{\Theta}]\right)_{cp}+4\left(\bm{\sigma}^T\bm{\sigma \Gamma} \textbf{F}[\bm{\Theta}^T\bm{\Theta}]\right)_{cp}\\
        &\left.+4\left(\bm{\sigma}^T\bm{\Theta}\textbf{F}[\bm{\Theta}^T\bm{\sigma\Gamma}]\right)_{cp}\right)\\
    \end{split}
\end{equation}

Let us now define different pieces of L,
\begin{equation}
    \label{eqn:l_pieces}
    \begin{split}
        L_1&=4c_0\mathrm{Tr}[\bm{\Theta} \textbf{G}\bm{\Theta}^T]+4\mathrm{Tr}[\bm{\Theta} \textbf{G}\bm{\Gamma}^T\bm{\sigma}^T] \\
        &+2c_0F[\bm{\Theta}^T\bm{\Theta}]\cdot\left(\bm{\Theta}^T\bm{\Theta}\right)+4\textbf{F}[\bm{\Theta}^T\bm{\Theta}]\cdot\left(\bm{\Theta}^T\bm{\sigma\Gamma}\right) \\
        L_2&=4c_0\mathrm{Tr}[\bm{\Theta}\textbf{G}\bm{\Gamma}^T\bm{\mu}^T]+4\mathrm{Tr}[\bm{\sigma\Gamma} \textbf{G}\bm{\Gamma}^T\bm{\mu}^T]\\
        &-2\mathrm{Tr}[\bm{\Theta}\textbf{G}\bm{\Theta}^T\bm{\sigma\mu}^T+\bm{\Theta} \textbf{G}\bm{\Theta}^T\bm{\mu\sigma}^T]+8\bm{\sigma}\cdot\bm{\mu}\mathrm{Tr}[\bm{\Theta} \textbf{G}\bm{\Theta}^T]\\
        &+4c_0\textbf{F}[\bm{\Theta}^T\bm{\Theta}]\cdot\left(\bm{\Theta}^T\bm{\mu\Gamma}\right)\\
        &+2\textbf{F}[\bm{\Theta}^T\bm{\Theta}]\cdot\left(\bm{\Gamma}^T\bm{\sigma}^T\bm{\mu\Gamma}+\bm{\Gamma}^T\bm{\mu}^T\bm{\sigma \Gamma}-\bm{\Theta}^T\bm{\mu\sigma}^T\bm{\Theta}-\bm{\Theta}^T\bm{\sigma\mu}^T\bm{\Theta}\right)\\
        &+4\left(\bm{\sigma}\cdot\bm{\mu}\right)\left(\textbf{F}[\bm{\Theta}^T\bm{\Theta}]\cdot\left(\bm{\Theta}^T\bm{\Theta}\right)\right)+4\textbf{F}[\bm{\Theta}^T\bm{\sigma\Gamma}]\cdot\left(\bm{\Theta}^T\bm{\sigma\Gamma}\right)\\
        L_3&=4N_2\mathrm{Tr}[\textbf{RG}\bm{\Theta}^T]+4c_0\mathrm{Tr}[\textbf{RG}\bm{\Gamma}^T\bm{\sigma}^T] \\
        &-2\mathrm{Tr}[\textbf{RG}\bm{\Theta}^T\bm{\sigma\sigma}^T+\bm{\Theta}\textbf{GR}^T\bm{\sigma\sigma}^T] \\
        &+2N_2\left(\textbf{F}[\textbf{R}^T\bm{\Theta}]+\textbf{F}[\textbf{R}^T\bm{\Theta}]^T\right)\cdot\left(\bm{\Theta}^T\bm{\Theta}\right)\\
        &+4c_0\left(\textbf{F}[\textbf{R}^T\bm{\Theta}]+\textbf{F}[\textbf{R}^T\bm{\Theta}]^T\right)\cdot\left(\bm{\Theta}^T\bm{\sigma\Gamma}\right) \\
        &+4c_0\textbf{F}[\bm{\Theta}^T\bm{\Theta}]\cdot\left(\textbf{R}^T\bm{\sigma\Gamma}\right) \\
        &+2\left(\textbf{F}[\textbf{R}^T\bm{\Theta}]+\textbf{F}[\textbf{R}^T\bm{\Theta}]^T\right)\cdot\textbf{A} \\
        &-2\textbf{F}[\bm{\Theta}^T\bm{\Theta}]\cdot\left(\textbf{R}^T\bm{\sigma\sigma}^T\bm{\Theta}+\left(\textbf{R}^T\bm{\sigma\sigma}^T\bm{\Theta}\right)^T\right) \\
        &+4\textbf{F}[\textbf{R}^T\bm{\sigma\Gamma}]\cdot\left(\bm{\Theta}^T\bm{\sigma\Gamma}\right) \\
        &+4c_0\mathrm{Tr}[\bm{\Theta}\textbf{G}\bm{\Phi}^T\bm{\sigma}^T]+2\mathrm{Tr}[\bm{\sigma\Gamma} \textbf{G}\bm{\Phi}^T\bm{\sigma}^T+\left(\bm{\sigma\Gamma}\textbf{G}\bm{\Phi}^T\bm{\sigma}^T\right)^T]\\
        &+4c_0\textbf{F}[\bm{\Theta}^T\bm{\Theta}]\cdot\left(\bm{\Theta}^T\bm{\sigma\Phi}\right) \\
        &+2\textbf{F}[\bm{\Theta}^T\bm{\Theta}]\cdot\left(\bm{\Phi}^T\bm{\sigma}^T\bm{\sigma\Gamma}+\left(\bm{\Phi}^T\bm{\sigma}^T\bm{\sigma
        \Gamma}\right)^T\right) \\
        &+4\textbf{F}[\bm{\Theta}^T\bm{\sigma\Gamma}]\cdot\left(\bm{\Theta}^T\bm{\sigma\Phi}\right)
    \end{split}
\end{equation}


The derivatives of $L_1$, $L_2$, and $L_3$ with respect to $c_0$ are,
\begin{equation}
    \label{eqn:l1234_der_c0}
    \begin{split}
        \frac{\partial L_1}{\partial c_0}&= 4\mathrm{Tr}[\bm{\Theta} \textbf{G}\bm{\Theta}^T]+2\textbf{F}[\bm{\Theta}^T\bm{\Theta}]\cdot\left(\bm{\Theta}^T\bm{\Theta}\right)\\
        \frac{\partial L_2}{\partial c_0}&=
        4\mathrm{Tr}[\bm{\Theta}\textbf{G}\bm{\Gamma}^T\bm{\mu}^T]+4F[\bm{\Theta}^T\bm{\Theta}]\cdot\left(\bm{\Theta}^T\bm{\mu\Gamma}\right) \\
        \frac{\partial L_3}{\partial c_0}&=
        8c_0\mathrm{Tr}[\textbf{RG}\bm{\Theta}^T]+4\mathrm{Tr}[\textbf{RG}\bm{\Gamma}^T\bm{\sigma}^T] \\
        &+4c_0\left(\textbf{F}[\textbf{R}^T\bm{\Theta}]+\textbf{F}[\textbf{R}^T\bm{\Theta}]^T\right)\cdot\left(\bm{\Theta}^T\bm{\sigma\Gamma}\right) \\
        &+4\textbf{F}[\bm{\Theta}^T\bm{\Theta}]\cdot\left(\textbf{R}^T\bm{\sigma\Gamma}\right) \\
        &+4\mathrm{Tr}[\bm{\Theta}\textbf{G}\bm{\Phi}^T\bm{\sigma}^T]+4\textbf{F}[\bm{\Theta}^T\bm{\Theta}]\cdot\left(\bm{\Theta}^T\bm{\sigma\Phi}\right)
    \end{split}
\end{equation}
in which no new Fock build is needed. 

The derivatives of $L_1$, $L_2$, and $L_3$ with respect to $\sigma_{jb}$ are,
\begin{gather}
    \label{eqn:l1234_der_sigma}
    \begin{split}
        \frac{\partial L_1}{\partial \sigma_{jb}}&=4\left(\bm{\Theta}\textbf{G}\bm{\Gamma}^T\right)_{jb}+4\left(\bm{\Theta} \textbf{F}[\bm{\Theta}^T\bm{\Theta}]\bm{\Gamma}^T\right)_{jb} \\
        \frac{\partial L_2}{\partial \sigma_{jb}}&=4\left(\bm{\Gamma}\textbf{G}\bm{\Gamma}^T\bm{\mu}^T\right)^T_{jb}-4\left(\bm{\Theta} \textbf{G}\bm{\Theta}^T\bm{\mu}\right)_{jb}+8\mathrm{Tr}[\bm{\Theta}\textbf{G}\bm{\Theta}^T]\mu_{jb} \\
        &+2\left(\bm{\mu\Gamma}\left(\textbf{F}[\bm{\Theta}^T\bm{\Theta}]+\textbf{F}[\bm{\Theta}^T\bm{\Theta}]^T\right)\bm{\Gamma}^T\right)_{jb} \\
        &-2\left(\bm{\Theta}\left(\textbf{F}[\bm{\Theta}^T\bm{\Theta}]+\textbf{F}[\bm{\Theta}^T\bm{\Theta}]^T\right)\bm{\Theta}^T\bm{\mu}\right)_{jb} \\
        &+4\textbf{F}[\bm{\Theta}^T\bm{\Theta}]\cdot\left(\bm{\Theta}^T\bm{\Theta}\right)\mu_{jb}+4\left(\bm{\Theta} \textbf{F}[\bm{\Theta}^T\bm{\mu\Gamma}]\bm{\Gamma}^T\right)_{jb} \\
        \frac{\partial L_3}{\partial \sigma_{jb}}&=4c_0\left(\textbf{RG}\bm{\Gamma}^T\right)_{jb}-4\left(\textbf{RG}
        \bm{\Theta}^T\bm{\sigma}\right)_{jb}+16\mathrm{Tr}[\textbf{RG}\bm{\Theta}^T]\sigma_{jb} \\
        &+4c_0\left(\bm{\Theta}\left(\textbf{F}[\textbf{R}^T\bm{\Theta}]+\textbf{F}[\textbf{R}^T\bm{\Theta}]^T\right)\bm{\Gamma}^T\right)_{jb} \\
        &+4\left(\bm{\sigma\Gamma}\left(\textbf{F}[\textbf{R}^T\bm{\Theta}]+\textbf{F}[\textbf{R}^T\bm{\Theta}]^T\right)\bm{\Gamma}^T\right)_{jb} \\
        &-4\left(\bm{\Theta}\left(\textbf{F}[\textbf{R}^T\bm{\Theta}]+\textbf{F}[\textbf{R}^T\bm{\Theta}]^T\right)\bm{\Theta}^T\bm{\sigma}\right)_{jb} \\
        &+4c_0\left(\textbf{RF}[\bm{\Theta}^T\bm{\Theta}]\bm{\Gamma}^T\right)_{jb} \\
        &-2\left(\textbf{R}\left(\textbf{F}[\bm{\Theta}^T\bm{\Theta}]+\textbf{F}[\bm{\Theta}^T\bm{\Theta}]^T\right)\bm{\Theta}^T\bm{\sigma}\right)_{jb} \\
        &-2\left(\bm{\Theta}\left(\textbf{F}[\bm{\Theta}^T\bm{\Theta}]+\textbf{F}[\bm{\Theta}^T\bm{\Theta}]^T\right)\textbf{R}^T\bm{\sigma}\right)_{jb} \\
        &+4\left(\textbf{RF}[\bm{\Theta}^T\bm{\sigma\Gamma}]\textbf{U}^T\right)_{jb}+4\left(\bm{\Theta}\textbf{F}[\textbf{R}^T\bm{\sigma\Gamma}]\bm{\Gamma}^T\right)_{jb} \\
        &+8F[\Theta^T\Theta]\cdot\left(R^T\Theta+\theta R^T\right)\sigma_{jb} \\
        &+4c_0\left(\Phi G\Theta^T\right)^T_{jb}+4\left(\Phi G\Gamma^T\sigma^T\right)^T_{jb}+4\left(\Gamma G\Phi^T\sigma^T\right)^T_{jb} \\
        &+4c_0\left(\Theta F[\Theta^T\Theta]\Phi^T\right)_{jb} \\
        &+2\left(\sigma \Phi\left(F[\Theta^T\Theta]+F[\Theta^T\Theta]^T\right)\Gamma^T\right)_{jb} \\
        &+2\left(\sigma \Gamma\left(F[\Theta^T\Theta]+F[\Theta^T\Theta]^T\right)\Phi^T\right)_{jb} \\
        &+4\left(\Theta\left(F[\Theta^T\sigma\Gamma]+F[\Theta^T\sigma\Gamma]^T\right)\Phi^T\right)_{jb} \\
        &+4\left(\Theta\left(F[\Theta^T\sigma\Phi]+F[\Theta^T\sigma\Phi]^T\right)\Gamma^T\right)_{jb} \\
    \end{split}
\end{gather}


The derivatives of $L_1$, $L_2$, and $L_3$ with respect to $\Theta_{kp}$ are,
\begin{equation}
    \label{eqn:l1234_der_ukp}
    \begin{split}
        \frac{\partial L_1}{\partial\Theta_{kp}}&=8c_0\left(\Theta G\right)_{kp}+4\left(\sigma \Gamma G\right)_{kp} \\
        &+8c_0\left(\Theta F[\Theta^T\Theta]\right)_{kp}+4\left(\sigma\Gamma F[\Theta^T\Theta]\right)_{kp} \\
        &+4\left(\Theta\left(F[\Theta^T\sigma\Gamma]+F[\Theta^T\sigma\Gamma]^T\right)\right)_{kp} \\
        \frac{\partial L_2}{\partial\Theta_{kp}}&=4c_0\left(\mu \Gamma G\right)_{kp}-4\left(\mu\sigma^T\Theta G\right)_{kp}+16\sigma\cdot\mu\left(\Theta G\right)_{kp}\\
        &+4c_0\left(\mu\Gamma F[\Theta^T\Theta]^T\right)_{kp} \\
        &+4c_0\left(\Theta\left(F[\Theta^T\mu\Gamma]+F[\Theta^T\mu\Gamma]^T\right)\right)_{kp} \\
        &-4\left(\Theta\left(F[\Theta^T\mu\sigma^T\Theta]+F[\Theta^T\mu\sigma^T\Theta]^T\right)\right)_{kp} \\
        &-2\left(\left(\mu\sigma^T\Theta+\sigma\mu^T\Theta\right)\left(F[\Theta^T\Theta]+F[\Theta^T\Theta]^T\right)\right)_{kp} \\
        &+8\sigma\cdot\mu\left(\Theta\left(F[\Theta^T\Theta]+F[\Theta^T\Theta]^T\right)\right)_{kp} \\
        &+4\left(\mu\Gamma F[\Theta^T\sigma\Gamma]^T\right)_{kp}+4\left(\sigma\Gamma F[\Theta^T\mu \Gamma]^T\right)_{kp} \\
        \frac{\partial L_3}{\partial \Theta_{kp}}&=4N_2\left(RG\right)_{kp}-4\left(\sigma\sigma^TRG\right)_{kp} \\
        &+2N_2\left(R\left(F[\Theta^T\Theta]+F[\Theta^T\Theta]^T\right)\right)_{kp} \\
        &+4N_2\left(\Theta\left(F[R^T\Theta]+F[R^T\Theta]^T\right)\right)_{kp} \\
        &+4c_0\left(R\left(F[\Theta^T\sigma\Gamma]+F[\Theta^T\sigma\Gamma]^T\right)\right)_{kp} \\
        &+4c_0\left(\sigma\Gamma\left(F[R^T\Theta]+F[R^T\Theta]^T\right)\right)_{kp} \\
        &+4c_0\left(\Theta\left(F[R^T\sigma\Gamma]+F[R^T\sigma\Gamma]^T\right)\right)_{kp} \\
        &+4c_0\left(\Theta\left(F[\Theta^T\sigma\Phi]+F[\Theta^T\sigma\Phi]^T\right)\right)_{kp} \\
        &+4c_0\left(\sigma\Phi F[\Theta^T\Theta]\right)_{kp}+2\left(R\left(F[A]+F[A]^T\right)\right)_{kp} \\
        &-4\left(\sigma\sigma^T\Theta\left(F[R^T\Theta]+F[R^T\Theta]^T\right)\right)_{kp} \\
        &+4\left(\Theta\left(F[B]+F[B]^T\right)\right)_{kp} \\
        &-2\left(\sigma\sigma^TR\left(F[\Theta^T\Theta]+F[\Theta^T\Theta]^T\right)\right)_{kp} \\
        &+4\left(\sigma\Gamma F[R^T\sigma\Gamma]^T\right)_{kp} \\
        &+4\left(\sigma\Phi F[\Theta^T\sigma\Gamma]^T\right)_{kp} \\
        &+4\left(\sigma\Gamma F[\Theta^T\sigma\Phi]^T\right)_{kp} \\
    \end{split}
\end{equation}

The derivatives of $L_1$, $L_2$, and $L_3$ with respect to $\Gamma_{cp}$ are,
\begin{equation}
    \label{eqn:l1234_der_Ucp}
    \begin{split}
        \frac{\partial L_1}{\partial \Gamma_{cp}}&=4\left(\sigma^T\Theta G\right)_{cp}+4\left(\sigma^T\Theta F[\Theta^T\Theta]\right)_{cp} \\
        \frac{\partial L_2}{\partial\Gamma_{cp}}&=4c_0\left(\mu \Gamma G\right)_{cp}-4\left(\mu\sigma^T\Theta G\right)_{cp}\\
        &+4\left(\sigma^T\mu \Gamma G\right)_{cp}+4c_0\left(\mu^T\Theta F[\Theta^T\Theta]\right)_{cp} \\
        &+2\left(\sigma^T\mu \Gamma\left(F[\Theta^T\Theta]+F[\Theta^T\Theta]^T\right)\right)_{cp} \\
        &+2\left(\mu^T\sigma \Gamma\left(F[\Theta^T\Theta]+F[\Theta^T\Theta]^T\right)\right)_{cp} \\
        &+4\left(\sigma^T\Theta F[\Theta^T\mu \Gamma]\right)_{cp}+4\left(\mu^T\Theta F[\Theta^T\sigma \Gamma]\right)_{cp} \\
        \frac{\partial L_3}{\partial \Gamma_{cp}}&=4c_0\left(\sigma^TRG\right)_{cp}+4\left(\sigma^T\sigma\Phi G\right)_{cp} \\
        &+4c_0\left(\sigma^T\Theta\left(F[R^T\Theta]+F[R^T\Theta]^T\right)\right)_{cp} \\
        &+4\left(\sigma^T\sigma \Gamma\left(F[R^T\Theta]+F[R^T\Theta]^T\right)\right)_{cp} \\
        &+4c_0\left(\sigma^TRF[\Theta^T\Theta]\right)_{cp} \\
        &+4\left(\sigma^TRF[\Theta^T\sigma\Gamma]\right)_{cp} \\
        &+4\left(\sigma^T\Theta F[R^T\sigma\Gamma]\right)_{cp} \\
        &+2\left(\sigma^T\sigma \Phi\left(F[\Theta^T\Theta]+F[\Theta^T\Theta]^T\right)\right)_{cp} \\
        &+4\left(\sigma^T\Theta F[\Theta^T\sigma\Phi]\right)_{cp} \\
    \end{split}
\end{equation}

\end{appendices}

\end{document}



\title[]{
Supplemental Materials: Excited State Mean-Field Theory without Automatic Differentiation
}

\author{Luning Zhao}
\affiliation{
Department of Chemistry, University of Washington, Seattle, Washington 98195, USA 
}
\author{Eric Neuscamman}
 \email{eneuscamman@berkeley.edu.}
\affiliation{
Department of Chemistry, University of California, Berkeley, California 94720, USA 
}
\affiliation{Chemical Sciences Division, Lawrence Berkeley National Laboratory, Berkeley, CA, 94720, USA}

\date{\today}

\maketitle

\section{Molecular Geometries}
\begin{figure}[h!]
\centering
\includegraphics[width=8.5cm,angle=0,scale=0.60]{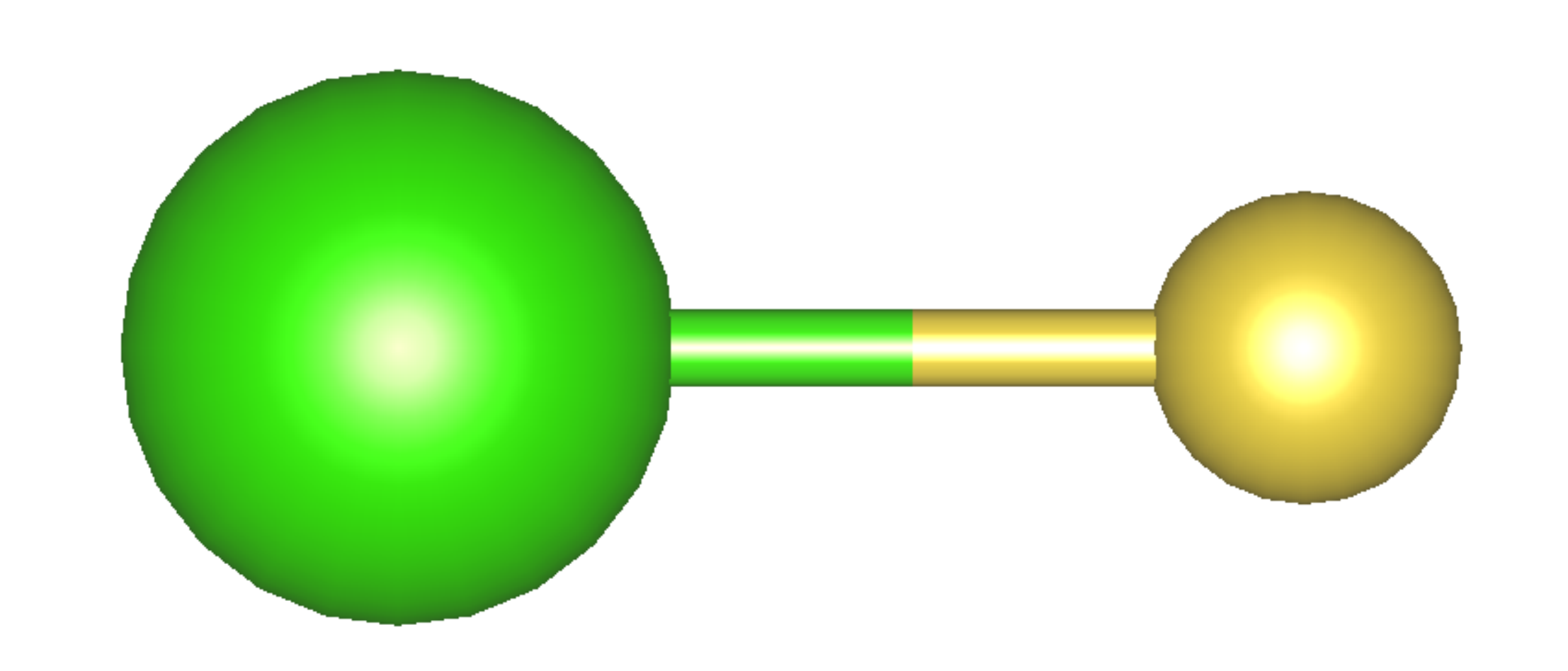}
\caption{Geometry of NaCl.}
\label{fig:nacl}
\end{figure}

\begin{figure}[h!]
\centering
\includegraphics[width=8.5cm,angle=0,scale=0.42]{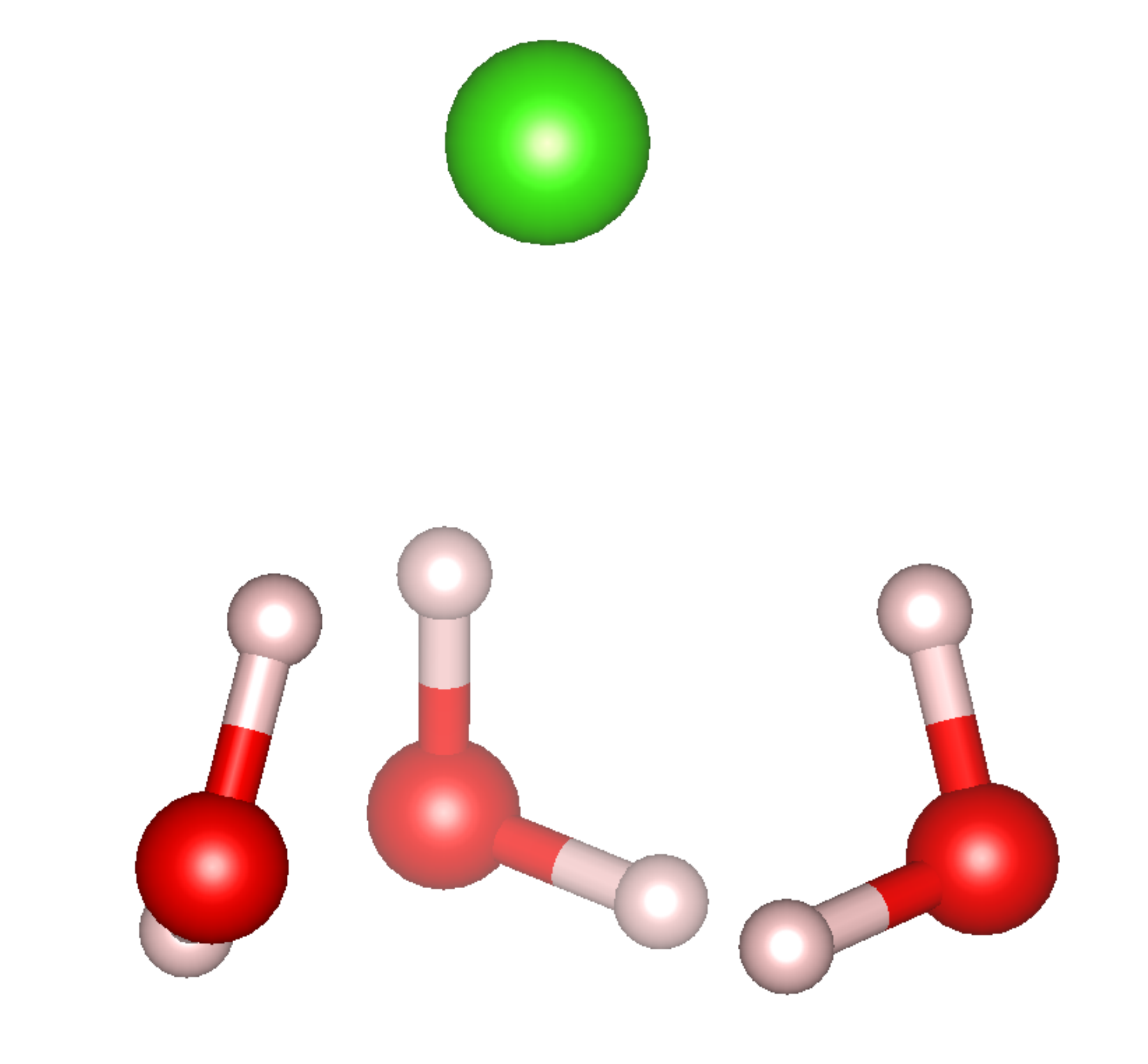}
\caption{Geometry of Cl$^-$-(H$_2$O)$_3$.}
\label{fig:cl_h2o_geo}
\end{figure}

\begin{figure}[h!]
\centering
\includegraphics[width=8.5cm,angle=0,scale=1.00]{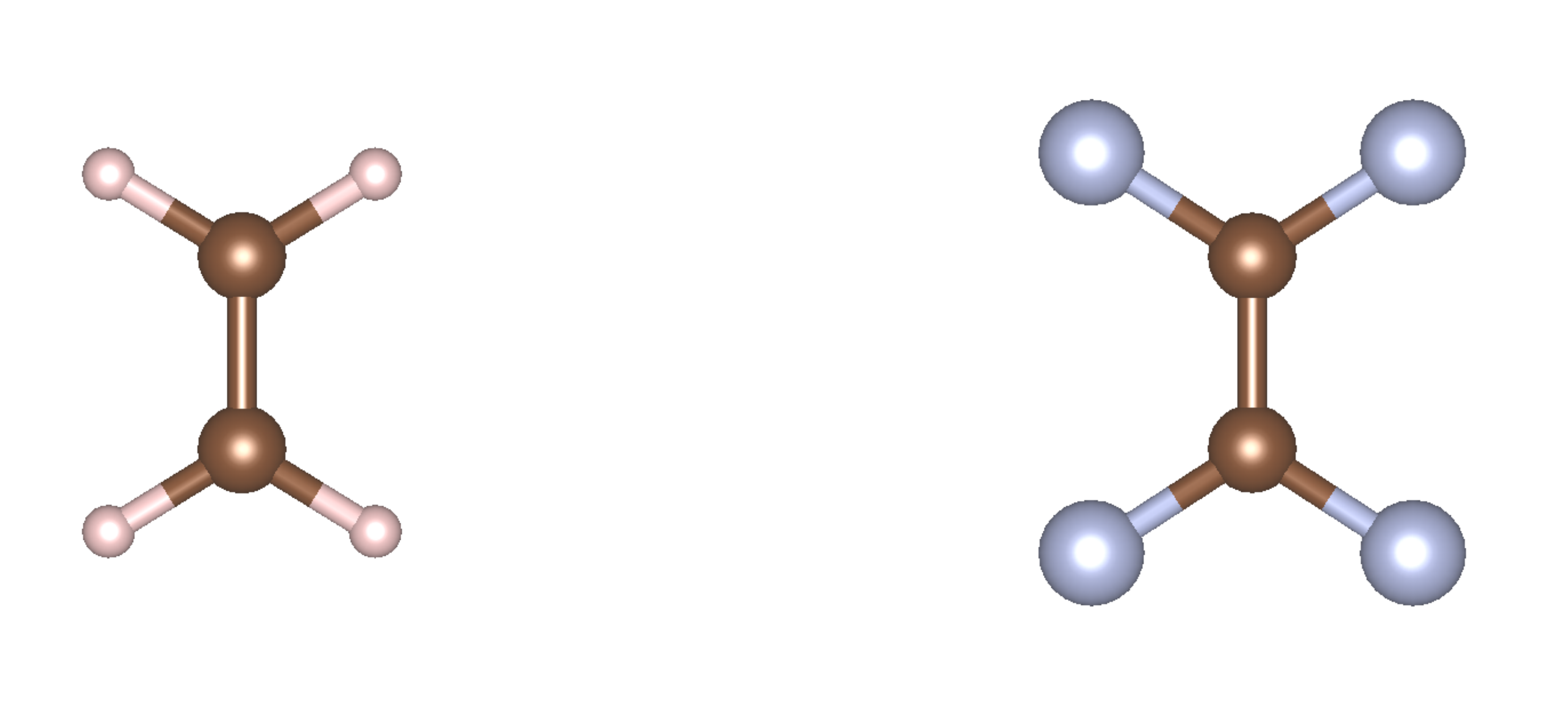}
\caption{Geometry of C$_2$H$_4$-C$_2$F$_4$.}
\label{fig:c2h4-c2f4_geo}
\end{figure}

\begin{figure}[h!]
\centering
\includegraphics[width=8.5cm,angle=0,scale=0.90]{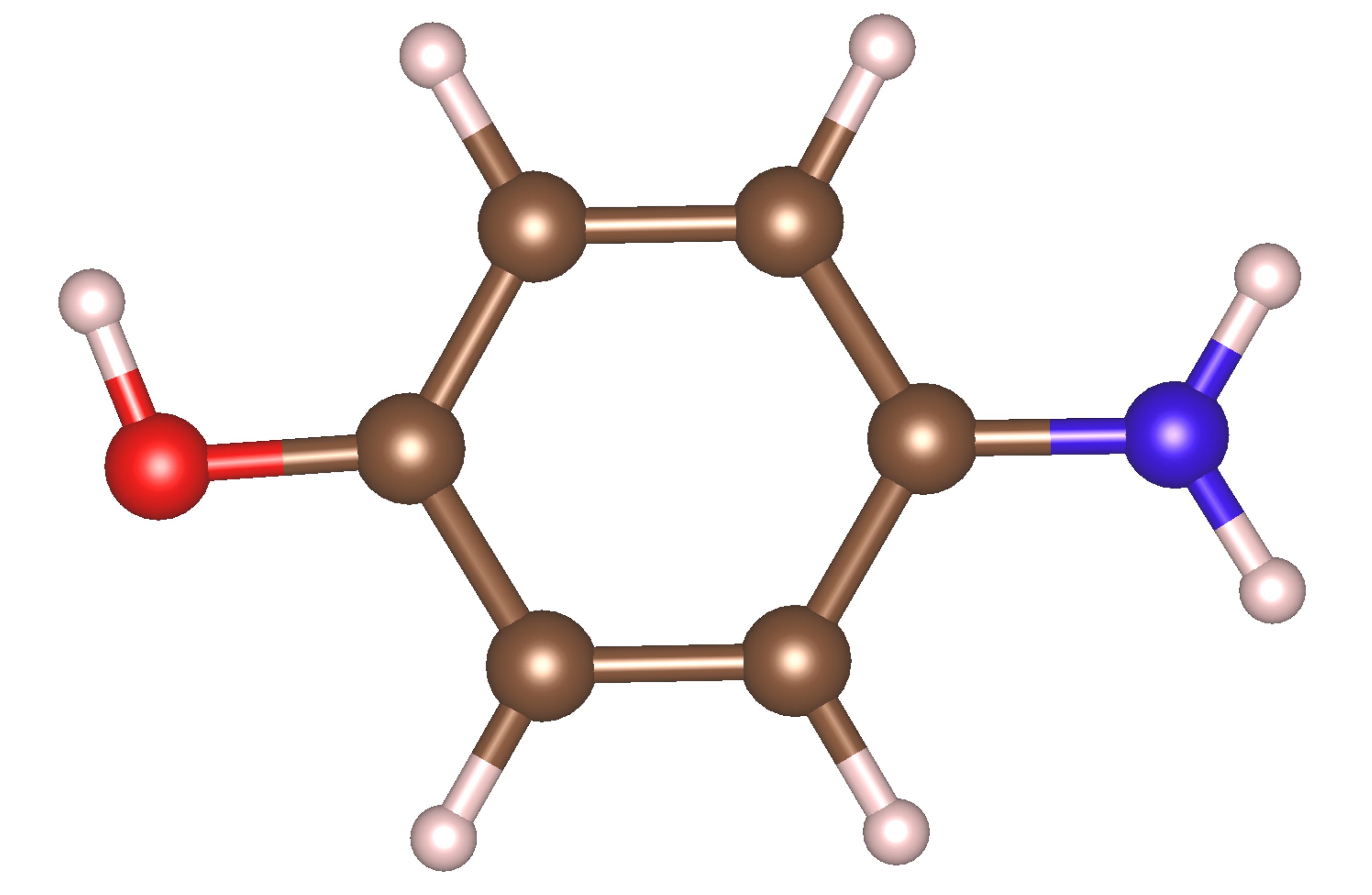}
\caption{Geometry of aminophenol.}
\label{fig:aminophenol_geo}
\end{figure}

The molecular structures of NaCl, Cl$^-$-(H$_2$O)$_3$, C$_2$H$_4$-C$_2$F$_4$, and aminophenol are 
show in Figure \ref{fig:nacl}, \ref{fig:cl_h2o_geo}, \ref{fig:c2h4-c2f4_geo}, 
and \ref{fig:aminophenol_geo}. We also include the Cartesian coordinates of these molecules, and the 
bond length used for NaCl is 2.36 $\mathring{A}$. 

The geometry of Cl-H$_2$O is, (units are in $\mathring{A}$)

Cl \hspace{0.2cm}  -0.848210 \hspace{0.2cm}   0.000015  \hspace{0.2cm}  0.074343

O  \hspace{0.38cm}   2.337083  \hspace{0.2cm}  0.000017  \hspace{0.2cm}  0.186525

H  \hspace{0.38cm}   1.693454  \hspace{0.2cm}  0.725249  \hspace{0.2cm}  0.163751

H  \hspace{0.38cm}   1.693562 \hspace{0.1cm}  -0.725306  \hspace{0.2cm}  0.163748

The geometry of NH$_3$-F$_2$ is,

N \hspace{0.3cm}  0.000000  \hspace{0.3cm} 0.000000 \hspace{0.2cm} 6.000000

H  \hspace{0.3cm} 0.000000  \hspace{0.3cm} 0.939731 \hspace{0.2cm} 5.611703

H \hspace{0.3cm}  0.813831 \hspace{0.2cm} -0.469865 \hspace{0.2cm} 5.611703

H \hspace{0.2cm} -0.813831 \hspace{0.2cm} -0.469865 \hspace{0.2cm}  5.611703

F \hspace{0.38cm}  0.000000 \hspace{0.3cm}  0.000000\hspace{0.2cm} 12.000000

F \hspace{0.38cm}  0.000000  \hspace{0.3cm} 0.000000\hspace{0.2cm} 13.430000

The geometry of Li-F system is, (units are in $\mathring{A}$)

Li   \hspace{0.15cm}     1.796944   \hspace{0.3cm}    0.402653  \hspace{0.2cm}     7.714191

O    \hspace{0.2cm}     0.913870   \hspace{0.3cm}    1.932310    \hspace{0.2cm}   6.845200

O     \hspace{0.2cm}    0.377075   \hspace{0.2cm}   -0.964195   \hspace{0.2cm}    7.698416

O     \hspace{0.2cm}    2.515306  \hspace{0.3cm}    0.844431    \hspace{0.2cm}   9.484318

O     \hspace{0.2cm}    3.091745   \hspace{0.2cm}   -0.174192   \hspace{0.2cm}    6.348637

H     \hspace{0.2cm}    3.349025     \hspace{0.3cm}  0.540540   \hspace{0.2cm}    9.879509

H     \hspace{0.2cm}    2.133055    \hspace{0.3cm}   1.467664    \hspace{0.15cm}   10.123912

H      \hspace{0.2cm}   3.887425    \hspace{0.2cm}  -0.730479   \hspace{0.2cm}    6.379252

H     \hspace{0.2cm}    3.192501    \hspace{0.3cm}   0.388996    \hspace{0.2cm}   5.562935

H      \hspace{0.2cm}   0.429815     \hspace{0.2cm} -1.696650    \hspace{0.2cm}   7.061571

H      \hspace{0.1cm}  -0.089280   \hspace{0.2cm}   -1.321908    \hspace{0.2cm}   8.472304

H     \hspace{0.2cm}    1.105006   \hspace{0.3cm}    2.884390    \hspace{0.2cm}   6.866262

H     \hspace{0.1cm}   -0.019955    \hspace{0.27cm}   1.852997    \hspace{0.2cm}   6.587612

F     \hspace{0.27cm}    0.243106    \hspace{0.2cm}  -0.058090    \hspace{0.05cm}  -0.170912

O     \hspace{0.2cm}    2.303862   \hspace{0.3cm}    1.382309    \hspace{0.2cm}   0.978656

O      \hspace{0.1cm}  -2.282249  \hspace{0.2cm} -0.646317 \hspace{0.05cm}  -1.078369

O      \hspace{0.2cm}   0.137241 \hspace{0.3cm}   2.682825  \hspace{0.05cm} -0.515020

O      \hspace{0.2cm}   2.626535  \hspace{0.2cm} -0.051411  \hspace{0.05cm} -1.623087

O     \hspace{0.1cm}   -1.783124 \hspace{0.3cm}   1.055466   \hspace{0.2cm} 1.206736

O      \hspace{0.2cm}  -1.154603  \hspace{0.2cm} -1.907153   \hspace{0.2cm} 1.277859

H      \hspace{0.2cm}   1.641197  \hspace{0.2cm}  0.687837 \hspace{0.2cm}   0.777891

H       \hspace{0.2cm}  1.835652 \hspace{0.2cm}   2.163873 \hspace{0.2cm}   0.627908

H      \hspace{0.2cm}  -1.334445  \hspace{0.2cm} -0.386089  \hspace{0.2cm} -1.061835

H      \hspace{0.2cm}  -2.258393  \hspace{0.2cm} -1.373147  \hspace{0.2cm} -0.426111

H      \hspace{0.2cm}  -0.447054  \hspace{0.2cm} -1.443480  \hspace{0.2cm}  0.773265

H       \hspace{0.2cm} -1.562261  \hspace{0.2cm} -1.153285  \hspace{0.2cm}  1.740562

H       \hspace{0.2cm}  2.944434  \hspace{0.2cm}  0.519587 \hspace{0.2cm}  -0.901516

H      \hspace{0.2cm}   1.727649  \hspace{0.2cm} -0.252872  \hspace{0.2cm} -1.297097

H      \hspace{0.2cm}  -0.614192  \hspace{0.2cm}  2.646830  \hspace{0.2cm}  0.104027

H      \hspace{0.2cm}   0.260635  \hspace{0.2cm}  1.727937 \hspace{0.2cm}  -0.704091

H       \hspace{0.2cm} -0.914261   \hspace{0.2cm} 0.722919 \hspace{0.2cm}   0.879951

H       \hspace{0.2cm} -2.352180 \hspace{0.2cm}   0.731683  \hspace{0.2cm}  0.478898

The geometry of Cl$^{-}$-(H$_2$O)$_3$ is, (units are in $\mathring{A}$)

Cl \hspace{0.2cm}   0.050329 \hspace{0.2cm}   0.035472  \hspace{0.2cm}  -2.015391

O  \hspace{0.38cm} -1.047843 \hspace{0.2cm}  -1.483881  \hspace{0.2cm}   0.617731

H  \hspace{0.38cm} -0.907505  \hspace{0.2cm} -1.210859  \hspace{0.2cm}  -0.310726

H  \hspace{0.38cm} -0.146825  \hspace{0.1cm} -1.432497  \hspace{0.2cm}   0.972306

O  \hspace{0.38cm} -0.843184  \hspace{0.2cm} 1.553755  \hspace{0.2cm}   0.686802

H  \hspace{0.38cm} -1.271587   \hspace{0.2cm} 0.728045  \hspace{0.2cm}   0.960088

H  \hspace{0.38cm} -0.626924   \hspace{0.2cm} 1.361459  \hspace{0.2cm}  -0.247573

O  \hspace{0.38cm}  1.675737   \hspace{0.2cm}-0.141942  \hspace{0.2cm}   0.774992

H  \hspace{0.38cm}  1.461043   \hspace{0.2cm}-0.169950  \hspace{0.2cm}  -0.178449

H  \hspace{0.38cm}  1.151415   \hspace{0.2cm} 0.618930  \hspace{0.2cm}   1.069156

The geometry of C$_2$H$_4$-C$_2$F$_4$ is, (units are in $\mathring{A}$)

C \hspace{0.2cm}   0.000000 \hspace{0.2cm}   0.000000  \hspace{0.2cm}  0.000000

C \hspace{0.2cm}   0.000000 \hspace{0.2cm}   1.336380  \hspace{0.2cm}  0.000000

H \hspace{0.2cm}   0.923269 \hspace{0.2cm}  -0.570117  \hspace{0.2cm}  0.000000

H \hspace{0.2cm}  -0.923269 \hspace{0.2cm}  -0.570117  \hspace{0.2cm}  0.000000

H \hspace{0.2cm}   0.923269 \hspace{0.2cm}   1.906497  \hspace{0.2cm}  0.000000

H \hspace{0.2cm}  -0.923269 \hspace{0.2cm}   1.906497  \hspace{0.2cm}  0.000000

C \hspace{0.2cm}   7.000000 \hspace{0.2cm}   0.004179  \hspace{0.2cm}  0.000000

C \hspace{0.2cm}   7.000000 \hspace{0.2cm}   1.332202  \hspace{0.2cm}  0.000000

F \hspace{0.2cm}   8.113928 \hspace{0.2cm}  -0.719141  \hspace{0.2cm}  0.000000

F \hspace{0.2cm}   5.886072 \hspace{0.2cm}  -0.719141  \hspace{0.2cm}  0.000000

F \hspace{0.2cm}   8.113928 \hspace{0.2cm}   2.055521  \hspace{0.2cm}  0.000000

F \hspace{0.2cm}   5.886072 \hspace{0.2cm}   2.055521  \hspace{0.2cm}  0.000000

The geometry of aminophenol is, (units are in $\mathring{A}$)

H \hspace{0.2cm}   1.247720 \hspace{0.2cm} -2.140105  \hspace{0.2cm}  0.000005

C \hspace{0.2cm}   0.694193 \hspace{0.2cm} -1.202493  \hspace{0.2cm}  0.000004

C \hspace{0.2cm}  -0.697351 \hspace{0.2cm} -1.230213  \hspace{0.2cm}  0.000000

H \hspace{0.2cm}  -1.235703 \hspace{0.2cm} -2.174088  \hspace{0.2cm} -0.000001

C \hspace{0.2cm}  -1.423742 \hspace{0.2cm} -0.037870  \hspace{0.2cm} -0.000001

O \hspace{0.2cm}  -2.800842 \hspace{0.2cm} -0.135154  \hspace{0.2cm} -0.000005

C \hspace{0.2cm}  -0.742494 \hspace{0.2cm}  1.180062  \hspace{0.2cm}  0.000001

H \hspace{0.2cm}  -1.295685 \hspace{0.2cm}  2.118976  \hspace{0.2cm} -0.000000

C \hspace{0.2cm}   0.653176 \hspace{0.2cm}  1.206486  \hspace{0.2cm}  0.000004

H \hspace{0.2cm}   1.170235 \hspace{0.2cm}  2.164577  \hspace{0.2cm}  0.000006

C \hspace{0.2cm}   1.395219 \hspace{0.2cm}  0.016120  \hspace{0.2cm}  0.000005

N \hspace{0.2cm}   2.778587 \hspace{0.2cm}  0.038060  \hspace{0.2cm}  0.000013

H \hspace{0.2cm}  -3.166110 \hspace{0.2cm}  0.766515  \hspace{0.2cm} -0.000007

H \hspace{0.2cm}   3.285260 \hspace{0.2cm}  0.907401  \hspace{0.2cm} -0.000018

H \hspace{0.2cm}   3.311516 \hspace{0.2cm} -0.815438  \hspace{0.2cm} -0.000017

\section{Isosurface Plots}
\begin{figure*}[h!]
    \centering
    \begin{subfigure}{0.3\textwidth}
        \centering
        \includegraphics[scale=0.30]{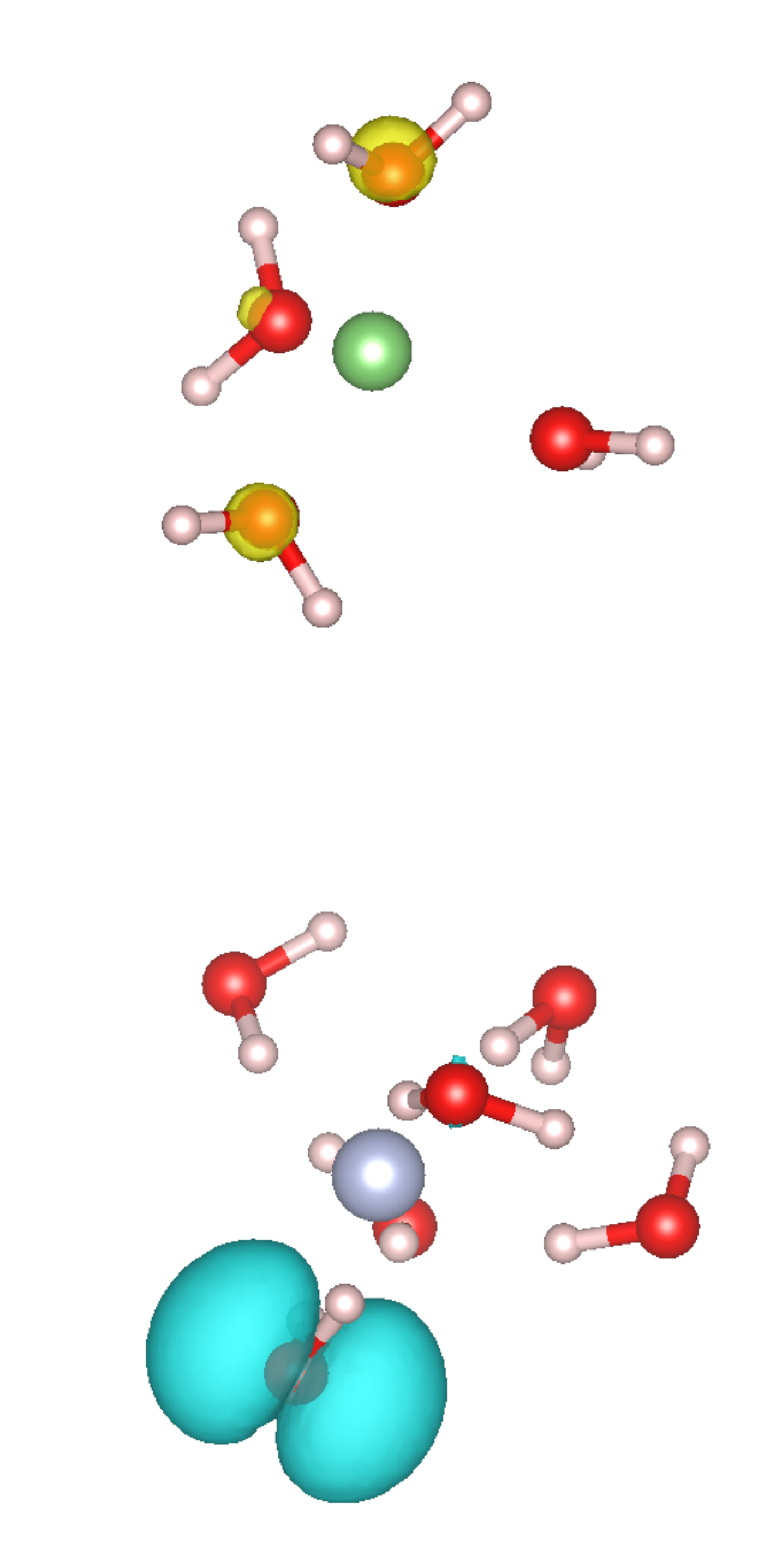}
        \caption{Surface Value=0.003}
        \label{fig:large_threshold}
    \end{subfigure}
    \qquad
    \begin{subfigure}{0.3\textwidth}
        \centering
        \includegraphics[scale=0.29]{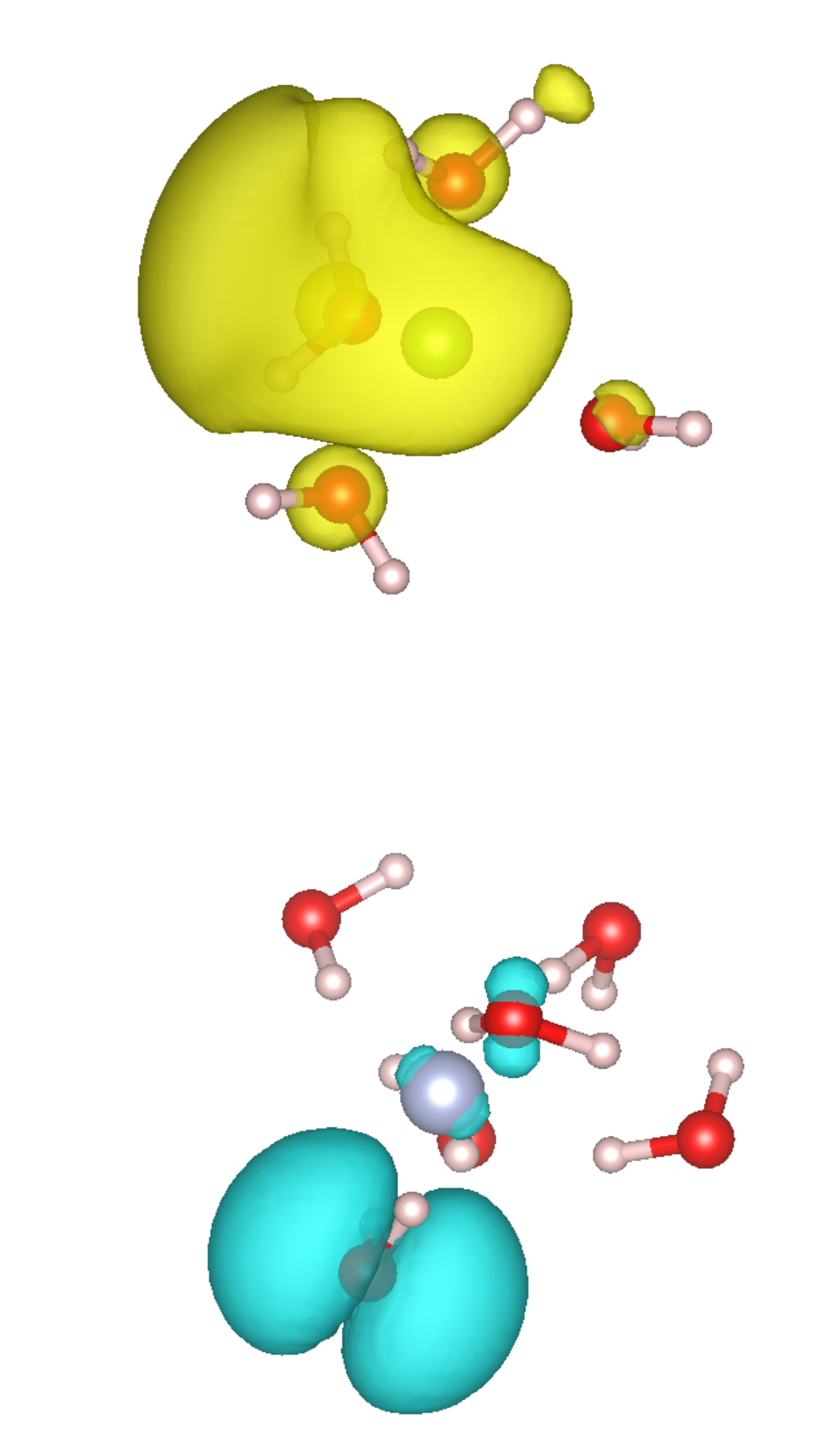}
        \caption{Surface Value=0.001}
        \label{fig:medium_threshold}
    \end{subfigure}
    \qquad
    \begin{subfigure}{0.3\textwidth}
        \centering
        \includegraphics[scale=0.29]{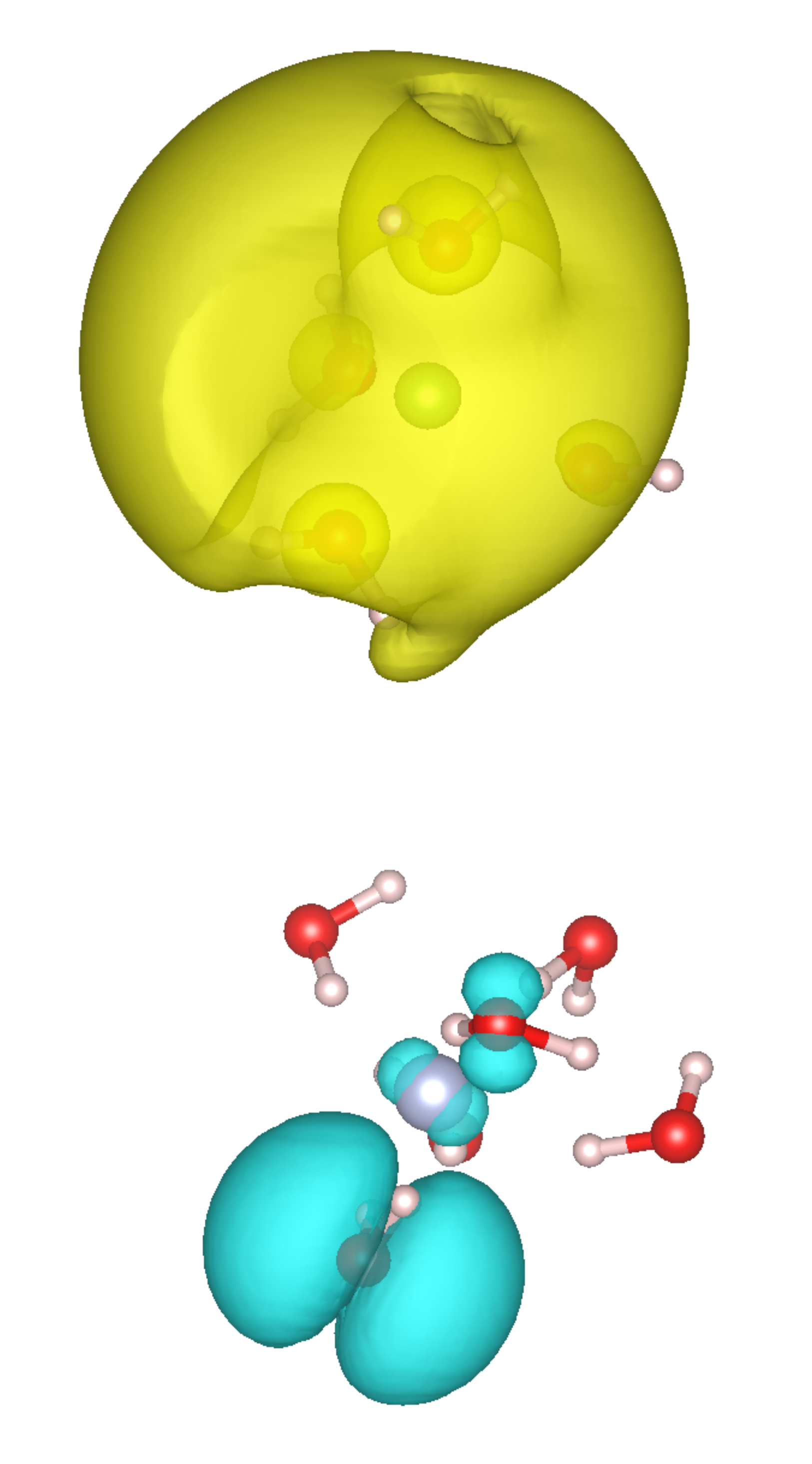}
        \caption{Surface Value=0.0005}
        \label{fig:small_threshold}
    \end{subfigure}
    \caption{CIS isosurface plots with different thresholds
             for the charge density changes following
             the charge transfer excitation
             that moves an electron from the lower F cluster to
             the upper Li cluster, with blue
             surfaces showing charge depletion relative to
             the ground state and yellow surfaces
             showing charge accumulation. 
            }
    \label{fig:cis_iso_diff_threshold}
\end{figure*}

\section{Overlap Between Ground and Excited State Wave Functions}
As ESMF does not enforce an orthogonality constraint, the optimized
excited state wave functions are not necessarily orthogonal to 
ground state. In this section we report the overlap between ground 
and excited state wave functions: $\left<\Psi_{\mathrm{gs}}|\Psi_{\mathrm{es}}\right>$. 
RHF determinant is used for ground 
state and ESMF wave function is used for excited state. In Table \ref{tab:gs_es_ovlp}
we show the  $\left<\Psi_{\mathrm{gs}}|\Psi_{\mathrm{es}}\right>$ for all the systems 
studied in this paper. We find the values are at most on the order 
of 10$^{-6}$ for all the systems tested. This shows that although the 
ESMF excited states are not rigorously orthogonal to RHF ground state, 
the amount of ground state contamination is negligible.

\begin{table*}[h!]
\caption{
  Overlap Between Ground and Excited State Wave Functions.
  \label{tab:gs_es_ovlp}
}
\begin{tabular}{c c}
\hline\hline
System & $\left<\Psi_{\mathrm{gs}}|\Psi_{\mathrm{es}}\right>$  \\
\hline
Cl$^-$-H$_2$O &   -2.90$\times$10$^{-6}$    \\
NH$_3$-F$_2$ &   1.59$\times$10$^{-7}$   \\
LiF-(H$_2$O)$_{10}$ &    2.27$\times$10$^{-6}$   \\
NaCl &  1.17$\times$10$^{-6}$  \\
Cl$^-$-(H$_2$O)$_3$ &  -2.20$\times$10$^{-6}$ \\
C$_2$H$_4$-C$_2$F$_4$ & 7.22$\times$10$^{-13}$   \\
aminophenol & 8.80$\times$10$^{-8}$ \\
\hline\hline
\vspace{2mm}
\end{tabular}
\end{table*}